\documentclass [12pt]{article}
\usepackage {amssymb}
\usepackage {amsmath}
\usepackage {graphicx}
\usepackage {longtable}

\begin{document}

\begin{flushleft}\footnotesize{ {\it Kybernetes: The International Journal of Systems
\& Cybernetics} {\bf 32}, No. 7/8, pp. 945-75 (2003) \\ {\rm (a
special issue on new theories about time and space, Eds.: L. Feng,
B. P. Gibson and Yi Lin)}}
\end{flushleft}

\section*{Scanning the structure of ill-known spaces: Part 1. \\
Founding principles about mathematical constitution of space}

\medskip
\begin{center}
{\bf Michel Bounias}$^{(1)}$ {\bf and Volodymyr
Krasnoholovets}$^{(2)}$
\end{center}

\begin{center} ($^{1}$) BioMathematics Unit(University/INRA, France, and
IHS, New York, USA), \\ Domain of Sagne-Soulier, 07470 Le Lac
d'Issarles, France            \\
($^{2}$) Institute of Physics, National Academy of Sciences, \\
Prospect Nauky 46,     UA-03028, Ky\"{\i}v, Ukraine
\end{center}

\textbf{Abstract}. Some necessary and sufficient conditions
allowing a previously unknown space to be explored through
scanning operators are reexamined with respect to measure theory.
Some generalized conceptions of distances and dimensionality
evaluation are proposed, together with their conditions of
validity and range of application to topological spaces. The
existence of a Boolean lattice with fractal properties originating
from nonwellfounded properties of the empty set is demonstrated.
This lattice provides a substrate with both discrete and
continuous properties, from which existence of physical universes
can be proved, up to the function of conscious perception.
Spacetime emerges as an ordered sequence of mappings of closed 3-D
Ponicar\'{e} sections of a topological 4-space provided by the
lattice, and the function of conscious perception is founded on
the same properties. Self-evaluation of a system is possible
against indecidability barriers through anticipatory mental
imaging occurring in biological brain systems; then our embedding
universe should be in principle accessible to knowledge. The
possibility of existence of spaces with fuzzy dimension or with
adjoined parts with decreasing dimensions is raised, together with
possible tools for their study.

The work presented here provides the introductory foundations supporting a
new theory of space whose physical predictions (suppressing the opposition
of quantum and relativistic approaches) and experimental proofs are
presented in details in Parts 2 and 3 of the study.

\bigskip \textbf{Key words:} Space structure; topological
distances; dimensionality assessment; Spacetime differential
element; Space lattice; objects origin.

\bigskip \textbf{PACS classification:} 02.10 Cz -- set theory;
02.40L Pc -- general topology; 03.65.Bz -- foundations, theory of
measurement, miscellaneous theories

\newpage
\section*{1. Introduction}

\hspace*{\parindent} Starting from perceptive aspects,
experimental sciences give rise to theoretical descriptions of
hidden features of the surrounding world. On the other hand, the
mathematical theory of demonstration teaches that any property of
a given object, from a canonical particle to the universe, must be
consistent with the characteristics of the corresponding embedding
space (Bounias, 2000a). In short, since what must be true in
abstract mathematical conditions should also be fulfilled upon
application to the observable world, whether the concept of
'reality' has a meaning or not. Conversely, since similar
predictions can infer when classical properties of Spacetime are
tested against various theories, e.g. an unbounded or a bounded
(non-Archimedean) algebra (Avinash and Rvachev, 2000), the most
general features should be accessible without previous assumptions
about peculiarities of the explored spaces.

Indeed, even the abstract branch of sciences, e.g. pure mathematics, show a
tendency to going to a form of "experimental research", essentially under
the pressure of some limitations that metamathematical considerations raise
about the fundamental questions of decidability (Chaitin, 1998-1999).
However, algorithmic information theory does not embrace the whole of
mathematics, and the theory of demonstration escapes the limitations of
arithmetical axiomatics, in particular through anticipatory processes
(Bounias, 2000b; 2001). To which extent the real world should obey just
arithmetics rules remains to be supported, and instead, the foundations of
the existence and functionality of a physical world have been shown to be
more widely provided by general topology (Bounias, 2000a). On the other
hand, how to explore a world considered as a system has been matter of
thorough investigation by Lin (1988; 1989).

The goal of the present work is to examine in depth some founding
mathematical conditions for a scientific scanning of a physical world to be
possible through definitely appropriate tools. In this respect, it is kept
in mind that human perceptions play a part, at least since humans (which
includes scientists) are self-conscious of their existence through their
perception of an outside world, while they believe in the existence of this
outside world because they perceive it. This "judge and party" antinomy will
be addressed in the present essay.

The first part of this essay is dealing with the notions of
measure and distances in a broad topological sense, including for
the assessment of the dimensionality of a space whose detached
pieces (i.e. the data collected through apparatus from a remote
world) are scatteredly displayed on the table of a scientist, in
view of a reconstruction of the original features. The existence
of an abstract lattice will be deduced and shown to stand for the
universe substrate (or "space"). A second part will focus on
specific features of this lattice and a confrontation of this
theoretical framework with the corresponding model of
Krasnoholovets and Ivanovsky (1993) about quantum to cosmic scales
of our observed universe will be presented, along with
experimental probes of both the theory and the model. A third part
will further present the structures predicted for elementary
particles whose existence derives from the described model, and
lead to a confrontation of the predictive performances of the
various theories in course.

\section*{2. Preliminaries}

\subsection*{2.1. About the concepts of measure and distances}

\hspace*{\parindent} Whatever the actual structure of our
observable spacetime, no system of measure can be operational if
it does not match with the properties of the measured objects.
Scanning an light-opaque world with a light beam, by ignorance of
the fundamental structure of what is explored, though a
caricatural example, illustrates the principle of a necessarily
failing device, those results would raise discrepancies in
knowledge of the studied world. An example is given by the recent
development of ultraviolet astronomy: when the sky is scanned
through ultraviolet instead of visible radiation, the resulting
extreme ultraviolet astrophysical picture of our surrounding
universe becomes different (Malina, 2000).

One of the problems faced by modeling unknown worlds could be
called \textit{"the syndrome of polynomial adjustment".} In
effect, given an experimental curve representing the behavior of a
system whose real mechanism is unknown, one can generally perform
a statistical adjustment with using a polynomial system like ${\rm
M}{\kern 1pt} {\kern 1pt} = \,\sum\nolimits_{\left( {\rm {i{\kern
1pt} = {\kern 1pt} 0{\kern 1pt} {\kern 1pt} \to N} }\right)}{\rm
{a_{i} \cdot \,} x^{i}}$. Then, using an apparatus adjusted to
test for the fitting of the N+l parameters to observational data
will require increasingly accurate adjustment, so as to
convincingly reflect the natural phenomenon within some
boundaries, while if the real equation is mathematically
incompatible with the polynomial, there will remain some
irreducible parts in the fitting attempts. This might well be what
occurs to the standard cosmological model and its 17 variables,
with its failure below some quantum scales (Arkani-Hamed et al.,
2000). Furthermore, Wu and Lin (2002) have demonstrated how the
approximations of solutions of equations describing nonlinear
systems mask the real structures of these systems. It may be that
what is tested in accelerators is a kind of self-evaluation of the
model, which poses a problem with respect to the indecidability of
computed systems as successively raised by Godel (1931), Church
(1934), Turing (1936), and more recently Chaitin, (1998-1999).
However, mathematical limits in computed systems can be overpassed
by the biological brain's system, due to its property of
self-decided anticipatory mental imaging (Bounias, 2001). It will
be shown here how this makes eventually possible a scanning of an
unknown universe by a part of itself represented by an internal
observer.

\subsubsection*{2.1.1. Measure}

\hspace*{\parindent} The concept of measure usually involves such
particular features as the existence of mappings and the
indexation of collections of subsets on natural integers.
Classically, a measure is a comparison of the measured object with
some unit taken as a standard (James and James, 1992). However,
sets or spaces and functions are measurable under various
conditions which are cross-connected. A mapping \textit{f} of a
set E into a topological space $\mathcal{T}$ is measurable if the
reciprocal image of a open of $\mathcal{T}$ by \textit{f}  is
measurable in E, while a set measure on E is a mapping
$\mathfrak{m}$ of a tribe B of sets of E in the interval $\left[
{0,\,\,\infty} \right]$,\textit{} exhibiting denumerable
additivity for any sequence of disjoint subsets (${\rm b_{{\kern
1pt} n}} $) of B, and denumerable finiteness, i.e. respectively:

  $$
\rm \mathfrak{m}\left( {{\mathop { \cup} \limits_{n = 0}^{\infty}
{\kern 1pt} {\kern 1pt} b_{n}} } \right) = \sum\limits_{n =
0}^{\infty} {\mathfrak{m}\left({\rm  {{\kern 1pt} {\kern 1pt}
b_{n}}  } \right)}
  \eqno(1.1)
  $$

  $$
{\rm \exists \,\,An,\,\quad An \in B,\quad \quad E{\kern 1pt}
{\kern 1pt} = {\kern 1pt} {\kern 1pt} {\kern 1pt} \cup
An,\;\;\,\,\;\forall {\kern 1pt} {\kern 1pt} \in \textbf {N},\;
\mathfrak{m} ( {An} ) }\ \  {\rm is \ finite}.
  \eqno(1.2)
  $$

Now, coming to the "unit used as a standard", this is the part
played by a gauge (J). Again, a gauge is a function defined on all
bounded sets of the considered space, usually having non-zero real
values, such that (Tricot, 1999): \\

\noindent (i) a singleton has measure naught: V x, J(\{x\}) = 0;

\noindent (ii) (J) is continued with respect to the Hausdorff
distance;

\noindent (iii) (J) is growing: E $\subset$ F $\Rightarrow$ J(E)
$\subset$ J(F);

\noindent (iv) (J) is linear: F(r$\cdot$E) : r$\cdot$J(E). \\

\noindent This implies that the concept of distance is defined:
usually, a diameter, a size, or a deviation are currently used,
and it should be pointed that such distances need to be applied on
totally ordered sets. Even the Caratheodory measure ($\mu^*$)
poses some conditions which again involve a common gauge to be
used: \\

\noindent (i) A $\subset$  B $\Rightarrow \mu$(A)$\leq \mu^*$(B);

\noindent (ii) for a sequence of subsets (Ei):
$\mu^*\subset$(Ri)${\kern 1pt} \leq {\kern 1pt}\Sigma {\kern 2pt}
\mu^*$(Ri);

\noindent (iii) $\angle$(A,B), A${\kern 1pt}\cap {\kern 1pt}{\kern
1pt}$B$=\O$: $\mu^*$(A${\kern 1pt} \cup {\kern
1pt}$B)$=\mu^*$(A)$+\mu^*$(B) (in consistency with (1.1));

\noindent (iv) $\mu^*$(E)$=\mu^*$(A${\kern 1pt}\cap {\kern
1pt}$E)$+\mu^*(\complement_{\rm E}{\kern 1pt}{\rm A}{\kern
1pt}\cap {\kern 1pt}$E). \\

\noindent The Jordan and Lebesgue measures involve respective
mappings (I) and ($\mathfrak{m}^*$) on spaces which must be
provided with $\cap$, $\cup$ and $\complement$. In spaces of the
$\mathbb{R}^{\rm n}$ type, tessellattion by balls are involved
(Bounias and Bonaly, 1996), which again demands a distance to be
available for the measure of diameters of intervals.

A set of measure naught has been defined by Borel (1912) first as a linear
set (E) such that, given a number (e) as small as needed, all points of E
can be contained in intervals whose sum is lower than (e).

\medskip
\noindent \textbf{Remark 2.1.1.} Applying Borel intervals imposes
that appropriate embedding spaces are available for allowing these
intervals to exist. This may appear as a not explicitly formulated
axiom, which might involve important consequences (see below).

\subsubsection*{2.1.2. A corollary on topological probabilities}

\hspace*{\parindent}
  Given a set measure (P) on a space E$=(($X,A),$\perp$), (A) a
tribe of parts of X, then (Chambadal, 1981): for a $\in$ A, P(a) =
Probability of (a) and P(X) = 1; for A,B in X, one has P(A u B) =
P(A) + P(B); for a sequence \{An\} of disjoint subspaces, one has:
$\lim_{{\kern 1pt}{\rm n}\rightarrow \infty}$ P(An) = 0.

A link can be noted with the Urysohn's theorem: let E, F be two
disjoint parts of a metric space W: there exists a continued
function {\it f} of W in the real interval [0,1] such that $f (\rm
{x})=0$ for any x $\in$ E, $\rm f(x)=1$ for any x $\in$ F, and
$0<f({\rm x})>1$ in all other cases. This has been shown to define
conditions providing a probabilistic form to a determined
structure holding for a deterministic event (Bonaly and Bounias,
1995). In effect, if W is the embedding space and S a particular
state of universe in W as recalled in section 3 below. Let C the
complementary of S in W, and x an object in the set of closed sets
in W. Let $\rm {I_{S}(x)}$ be an indicative function such that
${\rm I_{S}(x)}=1$ if x $\in$ S and $\rm I_S(x)=0$ if x $\in$ C.
Writing $\rm I_S(x)= P(X)$: x $\in$ S $\Rightarrow$ $\rm P(x)=1$;
x $\in$ X $\Rightarrow {\rm P(x)} = 0$ and $0 <{\rm  P(x)} < 1$ in
all other cases.

A probabilistic adjustment, as accurate as it can show, is thus not a proof
that a phenomenon is probabilistic in its essence.

\medskip

\subsubsection*{2.1.3. Distances}

\hspace*{\parindent} Following Borel, the length of an interval F
= [a,b] is:
  $$
{\rm L(F)= (b-a)-\sum_{n} L(C_ {{\kern 1pt}n})}
 \eqno(1.3)
  $$

\noindent where $\rm C_{{\kern 1pt} n}$ are the adjoined, i.e. the
open intervals inserted in the fundamental segment.

Such a distance is required in the Hausdorff distances of sets (E)
and (F): Let E(e) and F(e) the covers of E or F by balls B(x,e),
respectively for x $\in$ E or x $\in$ F,

 $$
{\rm  dist_{\kern 1pt H} (E,F) = inf} \{{\rm e: E \subset F(e)}
\wedge {\rm F \subset E(e)} \} \   \eqno(1.4{\kern 1pt}\rm a)
 $$
 $$
{\rm  dist_{\kern 1pt H} (E,F) = (x \subset E, \ y \subset F : inf
\ dist (x,y)) }
 \eqno(1.4{\kern 1pt}\rm b)
 $$

Since such a distance, as well as most of classical ones, is not
necessarily compatible with topological properties of the
concerned spaces, Borel provided an alternative definition of a
set with measure naught: the set (E) should be Vitali-covered by a
sequence of intervals (Un) such that: \ (i) each point of E
belongs to a infinite number of these intervals; \ (ii) the sum of
the diameters of these intervals is finite.

However, while the intervals can be replaced by topological balls, the
evaluation of their diameter still needs an appropriate general definition
of a distance.

 A more general approach (Weisstein, 1999b) involves a
path $\varphi$(x,y) such that $\varphi(0) ={\rm x}$ and
$\varphi(1) = {\rm y}$.

For the case of sets A and B in a partly ordered space, the
symmetric difference $\Delta {\rm (A,B) = \complement_{A {\kern
1pt}\cup{\kern 1pt} B}(A {\kern 1pt}\cap {\kern 1pt}B)}$ has been
proved to be a true distance also holding for more than two sets
(Bounias and Bonaly, 1996; Bounias, 1997-2000). However, if $\rm A
{\kern 1pt}\cup {\kern 1pt} B = \O$, this distance remains $\rm
\Delta = A {\kern 1pt}\cup{\kern 1pt} B$, regardless of the
situation of A and B within an embedding space E such that $\rm
(A,B){\kern 1pt} \subset {\kern 1pt} E$. A solution to this
problem will be derived below in terms of a separating distance
versus an intrinsic distance.

\subsection*{2.2. On the assessment of space dimensions}

\hspace*{\parindent} One important point is the following: in a
given set of which members structure is not previously known, a
major problem is the distinction between unordered N-uples and
ordered N-uples. This is essential for the assessment of the
actual dimension of a space.

\subsubsection*{2.2.1. Fractal to topological dimension}

\hspace*{\parindent} Given a fundamental segment (AB) and
intervals Li = [Ai, A(i+1)], a generator is composed of the union
of several such intervals: $\rm G = \cup_{{\kern 2pt} (i{\kern
1pt} \in {\kern 1pt}[1,{\kern 1pt }n])}(Li)$. Let the similarity
coefficients be defined for each interval by $\rm \varrho {\kern
1pt}i = dist (Ai, {\kern 1pt} A(i+1)) / dist (AB)$.

The similarity exponent of Bouligand is (e) such that for a
generator with n parts:

 $$
\rm \sum_{i{\kern 1pt}\in {\kern 1pt} [1,{\kern 1pt}n]}(\varrho
{\kern 1pt}i)^{{\kern 1pt }e} =1.
 \eqno(2.1)
 $$

When all intervals have (at least nearly) the same size, then the
various dimension approaches according to Bouligand, Minkowsky,
Hausdorff and Besicovitch are reflected in the resulting relation:

 $$
\rm  n \cdot (\varrho)^{{\kern 1pt }e}=l,
 \eqno(2.2{\kern 1pt}\rm a)
 $$

\noindent  that is:

 $$
\rm  e \approx Log {\kern 2pt} n / Log {\kern 2pt} \varrho .
 \eqno(2.2 \rm b)
 $$

When e is an integer, it reflects a topological dimension, since
this means that a fundamental space E can be tessellatted with an
entire number of identical balls B exhibiting a similarity with E,
upon coefficient $\varrho$.

\subsubsection*{2.2.2. Parts, ordered N-uples and simplexes}

\textbf{2.2.2.1. Parts.} A set is composed of members of which
some are themselves containing further members. In solving the
Russell and Burali-Forti paradoxes by making more accurate the
definition of a set, Mirimanoff (1917) classified members in
nuclei (i.e. singletons or "atoms") with no members inside
themselves, and parts containing members. Then, a set E is said
first kind (Ei) if it is isomorph to none of its members, and
second kind (En) if it is isomorph to at least one of its members.
Hence:

E = \{a,b,c, (d,e,f,)\} is first kind since $\rm X=(d,{\kern 1pt}
e,{\kern 1pt}f)$ is not isomorph to \{(a,b,c,(X)\}.

F = \{a, (b, (c, (Z))\} is second kind since by posing H=\{c,(Z)\}
and G = \{b, (H)\}, it appears that F = (a,(G)\} is isomorph to G
= \{b, (H)\},as well as to H, and eventually further to members of
H. Mirimanoff called a 'descent' a structure of the following
form, where (E) denotes a set or part of set, and (e) a nucleus:

 $$
\rm E^{(n)}= \{ e^{(n)}, (E^{n+1}) \}.
 \eqno(3.1)
 $$

A descent is finite if none of its parts is infinitely iterated.
Then, a second kind set is ordinary if its descent is finite and
extraordinary if its descent include some infinite part. One can
recognize a fractal feature in an extraordinary second kind set.

\medskip
\noindent \textbf{Remark 2.2.} A finite number (n) of iterations
provides a form of measure called the length of the descent. In
the above example of set F, assuming Z is not isomorph to F, the
sequence \{F, G, H\} is bijectively mapped on segment \{1, 2, 3\}
of natural integers.

\medskip
\noindent \textbf{2.2.2.1. ordered N-uples.} Let the members of
the above set E be ordered into a structure of the type of F, for
example: $\rm F^\prime = $\{a, (b, (c, (d,e,f)))\}.

The length of the descent of the exemplified $\rm
F^{\prime\prime}$ is $\rm L(F^\prime) = 3$, and the last part is
not isomorph to $\rm F^\prime$. Now, suppose that members (a, b,
c, ...) are similar, that is no particular structure nor essential
feature allow one to be distinguished from the other. Then,  $\rm
F^\prime$ could be written in the alternative form: $\rm
F^{\prime\prime} =$\{\{a\},\{a,\{a,...\}\}\}. This indicates the
availability of an order to hold on $\rm F^{\prime\prime}$.
Usually, a part (a,b) or \{a,b\} is not ordered until it can be
written in the form: (ab) = \{\{a\}, \{a,b\}\}. The nonempty part
of (ab) owns a lower boundary: \{a\}, and an upper boundary: (ab)
itself. Similarly:
 $$
\rm (abc) = \{\{a\},\ \{a,(a,b)\}, \ \{a, (a,b), \ (a,b,c)\}\}, \
etc.
 \eqno(3.2)
 $$

Stepping from a part (a,b,c,...) to a N-uple (abc) needs that
singletons are available in replicates. Two cases are met: if
members (a,b,c,...) are not identical, replicates can be found in
the set of parts of the set. Otherwise, if members are similar
singletons, then the set is just isomorphic with a segment of the
set of natural integers.

In both cases, for any pair of members (i.e. as subset members or
singletons), the Cartesian product will give a set of ordered
pairs. Repeating the operation in turn gives ordered N-uples. A
formal distinction of (xy), (x,y) and \{x,y\} will appear below.

\medskip
\noindent \textbf{2.2.2.2. Simplexes.} A simplex is the smaller
collection of points that allows the set to reach a maximum
dimension. In a general accepting, it should be noted that the
singletons of the set are called vertices and ordered N-uples are
(N-1)-faces A$^{\rm N-1}$. A set of (N+1) members can provide a
structure of dimension at most (N), that is: a connected "(N+l)-
object" has dimension $\rm d\leq N$. The number of 1-faces A$^{1}$
will be $\rm (N+1)N/2!$, the number of 2-faces A$^2$ will be $\rm
(N+1)\cdot N\cdot (N-1)/3!$, \ $\rm n(A^3) = (N+1)\cdot N\cdot
(N-1)\cdot (N-2)/4!$, \ and finally $\rm n(A^{k}) = (N+1)\cdot
N\cdot (N-1) ... (N-k+1)/(k+1)!$ (Banchoff, 1996).

One question emerging now with respect to the purpose of this
study is the following: given a set of N points, how to evaluate
the dimension of the space embedding these points?

\section*{3. Distances and dimensions revisited}

\subsection*{3.1. The relativity of a general form of measure and
distance}

\hspace*{\parindent} Our approach aims at searching for distances
that would be compatible with both the involved topologies and the
scanning of objects not yet known in the studied spaces. No such
configuration is believed to be an exception nor a general case.

\subsubsection*{3.1.1. General case of a not necessarily ordered
topological distance}

\textbf{Proposition 3.1.} A generalized distance between spaces
A,B within their common embedding space E is provided by the
intersection of a path-set $\varphi$(A,B) joining each member of A
to each member of B with the complementary of A and B in E, such
that: $\varphi$(A,B) is a continued sequence of a function $f$ of
a gauge (J) belonging to the ultrafilter of topologies on \{E, A,
B,...\}.

\medskip
\noindent \underline{Proof.} (i) A measure comes to a reality when
it maps a perceiving system (e.g. A) to a perceived one (e.g. B)
so that B is said measured by A within E. It has been formerly
shown that A and B should be topologically closed, since Jordan's
points are needed for the characterization of a path joining any
point of the interior of B to a point of the interior of A. The
nonempty intersection $\rm \underline{b}$ of the path with the
frontier $\partial{\kern 1pt}$B of B leads to the intersection
$\rm \underline{a}$ of the path with the frontier $\partial{\kern
1pt}$A of A, and a sequence of mappings $\rm u(\underline{a})$ of
$(\rm \underline{a})$ to a fixed point $\rm f(a)=a$ in A provides
the mathematical foundation of a mental image (Bounias and Bonaly,
1997) (Figure 1).

The path $\varphi$(A,B) is a set composed as follows: (i) $\rm
\varphi (A,B) = \cup_{{\kern 1pt}a\in A, {\kern 2pt} b \in B}
{\kern 2pt}\varphi (a,b)$, all defined in a sequence interval $\rm
[0, {\kern 2pt} f^{\kern 1pt n}(x)], \ x\in E$.

Then, for any closed D situated between A and B, $\rm f^{\kern 1pt
n }(m)$ intersects the frontiers of B, D, and A: thus, the
sequence $\rm f^{\kern 1pt n }$ has some of its points identified
with $\rm \underline{b}$, $\rm (\underline{di}, \ \underline{dj},
\ ... \ \in
\partial {\kern 1pt} D)$, and $\rm \underline{a}$. Therefore, the relative
distance of A and B in E, noted $\rm \Lambda_E(A,B)$ is contained
in $\varphi$(A,B):

 $$
\rm \Lambda_{E}(A,B) \subseteq\varphi(A,B)
 \eqno(4.1)
 $$

\begin{figure}
\begin{center}
\includegraphics[scale=2]{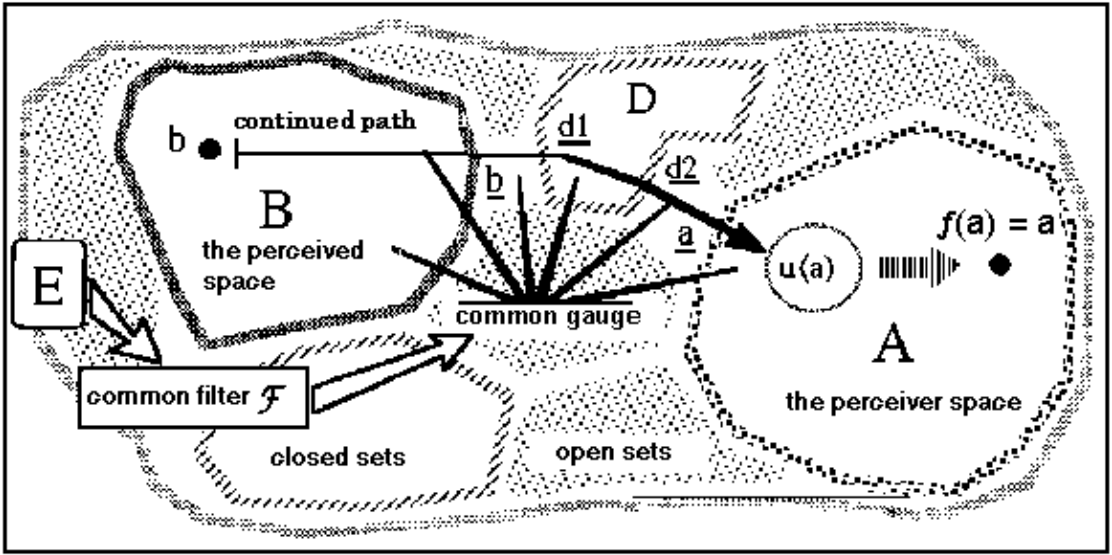}
\caption{\small{Some features of the measure of a distance between
closed sets (A,B) in an embedding space E. Space A is the
perceiver and space B is the perceived one. Points \underline{a},
\underline{b}, $\rm \underline{d\mbox{\scriptsize{1}}}$, $\rm
\underline{d\mbox{\scriptsize{2}}}$ are Jordan's points.}}
\label{1}
\end{center}
\end{figure}

Eventually one may have in the above case : min$\rm
\{\Lambda_{E}(A,B)\}$ = $\rm \{\underline{b}, \underline{di},
{\kern 2pt} \underline{dj}, \ ..., \ \underline{a}\}$, while in
all cases: inf$\{\rm \Lambda_E (A,B)\}$ = $\rm \{\underline{b}
\vee \underline{a}\}$. Denote by E$^\circ$ the interior of E,
then:

 $$
\rm min \{ \varphi(A,B) \cap E^\circ \} \ is \ a \ geodesic \ of \
space \ E \ connecting \ A \ to \ B \ \ \ \ \ \ \
 \eqno(4.2 \rm a)
 $$
 $$
\rm max  \{\varphi(A,B) \cap E^\circ  \cup \complement_{E^\circ}
(A{\kern 1pt}\cup {\kern 1pt} B)\} \ is \ tesselation \ of \ E \
out \ of \ A \ and \ B
 \eqno(4.2 \rm b)
 $$

It is noteworthy that relation (4.2a) refers to $\rm dim {\kern
1pt}\Lambda = dim {\kern 1pt} \varphi$, while in relation (4.2b)
the dimension of the probe is that of the scanned sets.

\bigskip \noindent (ii) Let $\rm J = f^{\kern 1pt}n(m)$, such that:
$\rm \varphi(0)=b$ and $\rm \varphi{\kern 1pt} f^{\kern 1pt
n}(m)=a$. Then, $\rm m {\kern 1pt} \in {\kern 1pt}
\mathcal{U}(E)$, with $\rm \mathcal{U}(E)$ the ultrafilter on
topologies of E.

Suppose that $\rm m {\kern 1pt} \notin {\kern 1pt}\mathcal{U}(E)$.
Then, there exists a filter $\rm \mathcal{F}_A$ holding on A and a
filter $\rm \mathcal{F}_B$  holding on B, such that $\rm
\mathcal{F}_A \neq \mathcal{F}_B $. Let $\rm x {\kern 1pt}\in
{\kern 1pt} \mathcal{F}_A$: there exists $\rm y {\kern 1pt}\in
{\kern 1pt} \mathcal{F}_B$ with $x\neq y$ and $\rm y{\kern
1pt}\notin {\kern 1pt}\mathcal{F}_A$.

Therefore, if $\rm x {\kern 1pt}\in {\kern 1pt}\varphi{\kern
1pt}f^{\kern 1pt n} (m)$, $\rm y \notin  {\kern 1pt}\varphi{\kern
1pt}f^{\kern 1pt n} (m)$ and the path-set does not measure B from
a perception by A. That $\rm m{\kern 1pt} \in {\kern
1pt}\mathcal{U}(E)$ is a necessary condition.

\bigskip
\noindent (iii) Let $(O)$ be a open set of E. Then, a member of
$\varphi$(A,B) joining B to A through \textit{O} meets no frontier
other than ${\kern 1pt}$B and $\partial{\kern 1pt}$A, and the
obtained $\rm \Lambda_E (A,B)$ ignores set \textit{O.} As a
consequence: only closed structures can be measured in a
topological space by a path founded on a sequence of Jordan's
points: this justifies and generalizes the Borel measure recalled
above. In contrast, if a closed set D contains closed subsets,
e.g. $\rm D^\prime \subset D $, then there is a member of
$\varphi$(A,B) that intersects $\rm D^\prime$. If in addition the
path is founded on $\rm f^{\kern 1pt n}(m)$, $\rm m{\kern 1pt} \in
{\kern 1pt}\mathcal{U}(E)$, there exists nonempty intersections of
the type $\rm \underline{d^\prime i}, \underline{d^\prime j}{\kern
1pt}\in {\kern 1pt} f^{{\kern 1pt} n}(m){\kern 1pt}\cap {\kern
1pt} \partial {\kern 1pt} D^\prime$. Therefore, $\rm \Lambda_E
(A,B)$ will include $\rm \underline{d^\prime i}$ and $\rm
\underline{d^\prime j}$ to the measured distance. This shows that
$\rm \varphi(A,B){\kern 1pt}\cap {\kern 1pt} E^\circ$ is a growing
function defined for any Jordan point, which is a characteristic
of a Gauge. In addition, the operator $\rm \Lambda_E(B,A)$ defined
by this way meets the characteristics of a Fr$\rm \acute{e}$chet
metrics, since the proximity of two points \underline{a} and
\underline{b} can be mapped into the set of natural integers and
even to the set of rational numbers: for that, it suffices that
two members $\rm \varphi \big( f^{{\kern 1pt}n}(x) \big),{\kern
1pt}f^{{\kern 1pt}n}(y) \big)$ are identified with an ordered pair
$\rm \{ \varphi \big( f^{{\kern 1pt}n}(x) \big),{\kern
1pt}\{\varphi \big( f^{{\kern 1pt}n}(x),{\kern 1pt} f^{{\kern
1pt}n}(y) \big) \}\}$.

\bigskip
\noindent (iv) Suppose that one path $\varphi$(a,b) meets an empty
space $\{\O \}$. Then a discontinuity occurs and there exists some
i such that $\rm \varphi \big( f^i(\underline{b})\big)$. If all
$\varphi$(A,B) meet $\{ \O \}$, then no distance is measurable. As
a corollary, for any singleton \{x\}, one has $\rm \varphi \big(
f(\{ x\}) \big) = \O$. The above properties meet two other
characteristics of a gauge.

\medskip \noindent
\textbf{Remarks 3.1.} Given closed sets \{A,B,C, ...\}= E, then a
path set $\rm \varphi(E,E\OE)$ exploring the distance of E to the
closure (\OE) of E meets only open subsets, so that $\rm \varphi
(E,E\OE) = \O$. This is consistent with a property of the
Hausdorff distance. Similarly, given A,B c E, one can tentatively
note:

 $$
\rm inf_{{\kern 1pt}x{\kern 1pt}\in {\kern 1pt} E} \{
\varphi(A,\{x\}) \cap E^\circ \} \ \mapsto \ dist_{{\kern 1pt}
Hausdorff}(\{x\},A)
  \eqno(4.3 \rm a)
 $$

\noindent as reported by Choquet (1984) or Tricot, 1999b);

 $$
\rm sup {\kern 1pt}\varphi \{(A,B) \cap E^\circ \} \  \mapsto  \
dist_{{\kern 1pt}Hausdorff}(A,B) \ \ \ \ \ \  \eqno(4.3 \rm b)
 $$
as reported by Tricot (1999b);

 $$
\rm inf_{{\kern 2pt}a \in A, {\kern 1pt}b \in B{\kern 2pt}}
\varphi \{(a,b) \cap E^\circ \} \ \mapsto \ dist (A \wedge B) \ \
\ \ \ \ \ \   \eqno(4.3 \rm c)
 $$
as reported by Choquet (1984) for a (E,d) metric space;

 $$
\rm max\{ (A, B) \subset E | \Lambda_{E}(A,B) \} \   \mapsto  \
diam_{{\kern 1pt}Hausdorff}(E)
 \eqno(4.3 \rm d)
 $$
in all cases.

\bigskip
\noindent (v)  $\rm \Lambda_E(A,B) = \Lambda_E(B,A)$ and $\rm
\Lambda_{E}(\{x,y\}) =\O$ $\rm \Leftrightarrow x=y$. If the
triangular inequality condition is fulfilled, then $\rm
\Lambda_{E}(B,A)$ will meet all of the properties of a
mathematical distance.

$\rm \Lambda(A,B){\kern 1pt} \cup {\kern 1pt} \Lambda_E(B,C)$ may
contain members of $\rm \Lambda_E(A,C)$ since the latter are
contained in neighborhoods of A,B,C. Thus:

 $$
\rm \Lambda_{E}(A,C)= \{ \exists{\kern 1pt} B, (A {\kern 1pt} \cap
{\kern 1pt} B \neq  \O, \ C {\kern 1pt}\cap {\kern 1pt} B \neq
\O), \ \Lambda_E(A,C) \subseteq \Lambda_E (A,B)\cup \Lambda_E
(B,C) \}.
 \eqno(4.3 \rm e)
 $$

This completes the proof of Proposition 2.1.

\subsubsection*{3.1.2. The particular case of a totally ordered
space}

\hspace*{\parindent} Let A and B be disjoint segments in space E.
Let E be ordered by the classical relations:

 $$
\rm A \subset B \ \Leftrightarrow  \ A \prec B, \  \quad \quad \ \
\
\
 \eqno(5.1 \rm a)
 $$
 $$
\rm (A,B) \subset \  \Leftrightarrow \ E \succ A, \ E \succ B.
 \eqno(5.1 \rm b)
 $$

Then, E is totally ordered if any segment owns a infinum and a
supremum. Therefore, a distance (d) between A and B is represented
as illustrated by Figure 2 by the following relation:

 $$
\rm d(A,B) \subseteq dist (inf {\kern 1pt} A, inf {\kern 1pt} B)
\cap {\kern 1pt} dist (sup{\kern 1pt} A,{\kern 1pt}sup {\kern 1pt}
B)
 \eqno (5.2)
 $$
with the distance evaluated through either classical forms or even
the set-distance $\Delta$(A,B) which will be revisited below.

\subsubsection*{3.1.3. The case of topological spaces}

\textbf{Proposition 3.2.} A space can be subdivided in two main
classes: objects and distances.

The set-distance is the symmetric difference between sets: it has
been proved that it owns all the properties of a true distance
(Bounias and Bonaly, 1996) and that it can be extended to
manifolds of sets (Bounias, 1997). In a topologically closed
space, these distances are the open complementary of closed
intersections called the "instances". Since the intersection of
closed sets is closed and the intersection of sets with nonequal
dimensions is always closed (Bounias and Bonaly, 1994), the
instances rather stands for closed structures. Since the latter
have been shown to reflect physical-like properties, they denote
objects. Then, the distances as being their complementaries will
constitute the alternative class: thus, a physical-like space may
be globally subdivided into objects and distances as full
components. This coarse classification will be further refined in
Part 2.

The properties of the set-distance allows an important theorem to be now
stated.

\medskip
\noindent \textbf{Theorem 3.1.} Any topological space is
metrizable as provided with the set-distance ($\Delta$) as a
natural metrics. All topological spaces are kinds of metric spaces
called "delta-metric spaces".

\medskip \noindent
\underline{Proof.} Conditions for a space X (generally belonging
to the set of parts of a space W) to be a topological space are
three folds (Bourbaki, 1990): \ (i) Any union of sets belonging to
X belongs to X. If A and B belong to X, then $\rm \Delta {\kern
1pt} (A,B)=\complement_{A \cup B}(A \cap B) \subseteq (A\cup B)
\subset X$; \ (ii) Any finite intersection of sets belonging to X
belongs to X. Let (A,B) $\in$ X. Since $\rm max {\kern 1pt} \Delta
(A,B)=(A\cup B)$ and $\rm min {\kern 1pt} \Delta (A,B)=\O$, and
that $\rm \O \in X$ (Schwartz, 1991), then necessarily $\rm (A,B)
\in X$. The symmetric distance fulfills the triangular inequality,
including in its generalized form, it is empty if $\rm A=B=...$,
and it is always positive otherwise. It is symmetric for two sets
and commutative for more than two sets. Its norm is provided by
the following relation : $\rm ||\Delta (A)|| = \Delta (A,\O)$.

Therefore, any topology provides the set-distance which can be called a
topological distance and a topological space is always provided with a self
mapping of any of its parts into any one metrics: thus any topological space
is metrizable.

Reciprocally, given the set-distance, since it is constructed on the
complementary of the intersection of sets in their union, it is compatible
with existence of a topology. Thus, a topological space is always a
"delta-metric" space.

\medskip
\noindent \textbf{Remark 3.2.} Distance $\Delta$(A,B) is a kind of
an intrinsic case [$\rm \Lambda_{(A,B)}$(A,B)] of $\rm
\Lambda_E(A,B)$ while $\rm \Lambda_E(A,B)$ is called a
\textsf{"separating distance"}. The separating distance also
stands for a topological metrics. Hence, if a physical space is a
topological space, it will always be measurable.

\begin{figure}
\begin{center}
\includegraphics[scale=1.2]{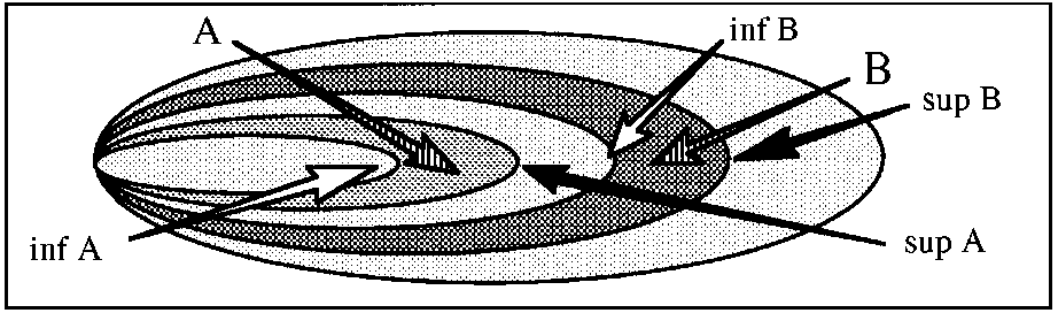}
\caption{\small{The distance of two disjoint segments in a totally
ordered space.}} \label{2}
\end{center}
\end{figure}

\subsection*{3.2. A corollary about intrinsic versus separating
distances}

\subsubsection*{\textbf{3.2.1. Introduction}}

\hspace*{\parindent} The previously proposed set distance as the
symmetric difference between two or more sets is independent of
any embedding space. It should thus be considered as an intrinsic
one. However, the measure based on a path provided with a gauge
pertaining to the common filter on A,B in E seeks for an
identification of what is between A and B within E. Thus, a
particular application can be raised :

\subsubsection*{\textbf{3.2.2. Results}}

\textbf{Proposition 3.2.} Let spaces (A,B) $\subset$ E. Then a
measure of the separating distance of A and B is defined if there
exists a space X with nonempty intersections with E, A, B, such
that, X belongs to the same filter $\mathcal{F}$ as A and B, and:

 $$
\rm \Lambda_{E}(A,B): E \cap \{\Delta(A,X) \cap \Delta(B,X)\}
 \eqno(6)
 $$

\medskip\noindent
\underline{Preliminary proof.} Since filter $\mathcal{F}$ holds on
E, and A,B$\in \mathcal{F}$ the three properties of a filter state
the following (Bourbaki, 1990b, 1.36):      \\

\noindent (i) {\kern 1pt} $\rm X \in \mathcal {F}$ implies X must
contain a set $\rm G \in \mathcal{F}$;

\noindent (ii) {\kern 1pt} Since any finite intersection of sets
of $\mathcal {F}$ must belong to $\mathcal {F}$, one has:
  $$
  \rm G \in \{ Ai, {\kern 1pt} Aj \in \mathcal{F}, \
i\neq j : \ Ai \cap Aj \};
  $$

\noindent (iii) {\kern 1pt} The empty part of X does not belong to
$\mathcal{F}$. Therefore since $\rm G \in X$ and $\rm A \cap
B=\O$, then there must be $\rm G \in A \cap X$ and $\rm G \in B
\cap X$, $\rm G \neq \O$, which proves (4.1).      \\

\noindent Hence, this example (Figure 3) further provides evidence
that a definition is just a particular configuration of the
intersection of two spaces of magmas of which one is the reasoning
system (eventually a logic) and the other one is a probationary
space.

\begin{figure}
\begin{center}
\includegraphics[scale=1.4]{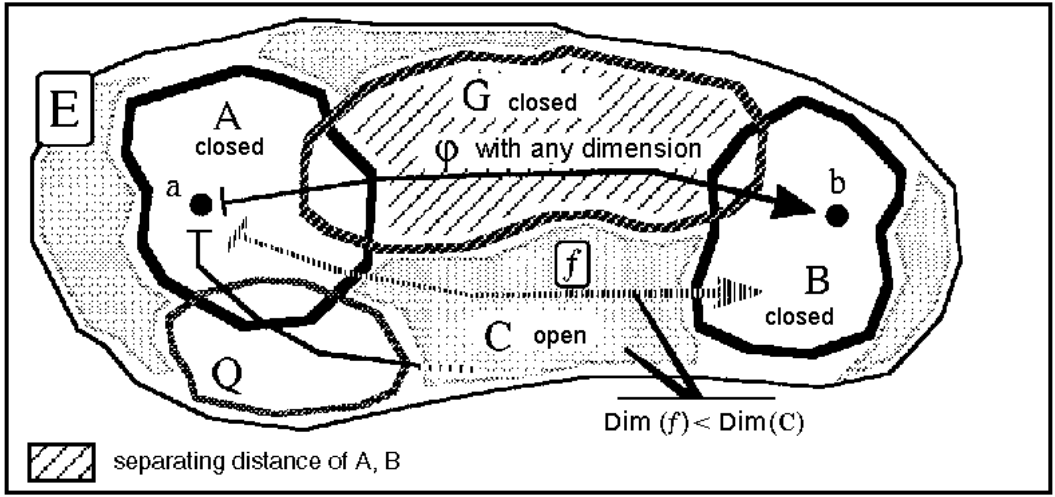}
\caption{\small{A path scanning the separating part between a
closed space A and another closed space B must own a nonempty
intersection with the objects situated between A and B. Sets A and
G and sets B and G own their symmetric difference as an intrinsic
set-distance. The intersection of these two measures is the
separating distance.  \newline \indent  \qquad {\bf Note:} A set Q
having empty intersection with either A or B (here with B) does
not provide a separating distance since it does not necessarily
fulfill the embedding space with connectivity. In this case, $\rm
\Delta(A,{\kern 1pt}B,{\kern 1pt}Q)= \Delta (A,{\kern 1pt}Q){\kern
1pt} \cup {\kern 1pt}\Delta(B,{\kern 1pt}{\O}).$ }} \label{3}
\end{center}
\end{figure}

\subsubsection*{3.2.3. Particular case: measuring open
sets}

\hspace*{\parindent} As pointed above, even a continued path
cannot in general scan a open component of the separating distance
between two sets, since a path has in general no closed
intersection with a open with same dimension. This is consistent
with the exclusion of open adjoined intervals in the Borel
measure. Hence, since a primary topology is a topology of open
sets, a primary topological space cannot be a physically
measurable space.

However, an intersection of a closed (C) with a path ($\varphi$)
having a nonequal dimension than (C) owns a closed intersection
with C provided this intersection is nonempty. This implies that
the general conditions of filter membership is fulfilled (Figure
4).

\begin{figure}
\begin{center}
\includegraphics[scale=1.4]{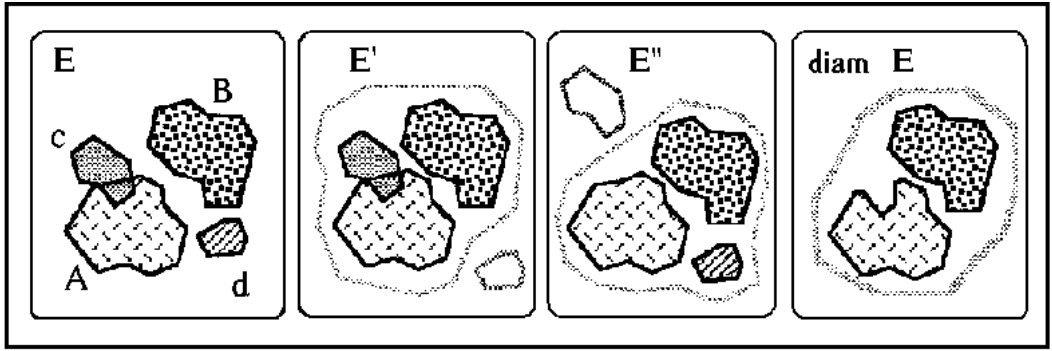}
\caption{\small{A preliminary approach of the diameter of an
abstract set in a partly ordered space: this examples allows the
sup(diam{\kern 1pt}E) as included in the complementary of c and d
in E. }} \label{4}
\end{center}
\end{figure}

\medskip
\noindent \textbf{Remark.} An open 3-D universe would not be
scanned by a 3-D probe. But in a closed Poincar\'{e} section, the
topologies are distributed into closed parts and their
complementaries as open subparts. Therefore, there may be open
parts in our universe that would not be detectable by 3-D probes.
This problem might be linked with the still pending problem of the
missing dark matter (see also Arkani-Hamed, 2000).

\subsubsection*{3.2.4. An alternative perspective}

\hspace*{\parindent} Owing to the case in which there exists no
intersection of spaces A and B with one of the other spaces
contained in E, in order to define a common surrounding of A and B
in an appropriate region of E, the following proposition would
hold:

\medskip
\noindent\textbf{Proposition 3.3.} A "surrounding distance" of
spaces A and B in an embedding space E is given by the
complementary of A and B in the interior of their common closure.
This distance is noted $\rm @_{E}(A,B)$:

 $$
\rm @_{E}(A,B) = \big[ \{ \overline{(A,B)^\circ} \}_{E} \backslash
{\kern 1pt}(A \cup B) \big] \supset \{ \overline{A}\} \cup \{
\overline {B}\}.
 \eqno(7)
 $$

\medskip
\noindent \textbf{Corollary 3.3.1.} A condition for the
surrounding distance to be non-zero, is that A and B must be dense
in E.

The closure of $\rm A \cup B$ is different from, and contains, the
union of the respective closures of A and B. This important
property clearly delimitates a region of space E where any object
may have to be scanned through a common gauge in order to allow a
measure like $\rm \Lambda _E(A,B)$ to be allowed. Hence, the
concept of surrounding distance is more general since it induces
that of separating distance and belongs to a coarser topological
filter.

\subsection*{3.3. About dimensionality studies}

\hspace*{\parindent} A collection of scientific observation
through experimental devices produce images of some reality, and
these images are further mapped into mental images into the
experimentalist's brain (Bounias, 2000a). The information from the
explored space thus stands like parts of an apparatus being spread
on the worker's table, in view of a further reconstitution of the
original object. We propose to call this situation an
informational display, likely composed of elements with dimension
lower than or equal to the dimension of the real object.

The next sections will thus deal with this particular problem.

\subsubsection*{3.3.1. Analysis of unordered versus ordered pairs in
an abstract set}

\hspace*{\parindent} A robust definition of a ordered N-uple is
given by the following:

\medskip
\noindent \textbf{Lemma 3.1} An expression noted (abc...z) is a
ordered N-uple iff:
 $$
\rm (abc...z) = (x_1{\kern 1pt} x_2 {\kern 1pt} x_3 {\kern 1pt}
... {\kern 1pt} x_n) \ \Leftrightarrow \  a=x_1, {\kern 2pt}
b=x_2, {\kern 2pt} c=x_3,{\kern 2pt} ..., {\kern 2pt} z=x_n.
 \eqno(8.1 \rm a)
 $$

In the construction of the set N of natural integers, Von Neumann
provided an equipotent form using replicates of \O:
  $$
0=\O, \ 1=\{\O \},{\kern 2pt} 2=\{\O,\{\O\}\},{\kern 2pt}3=\{\O,
\{\O\}, \{\O,\{\O\}\}\},
  $$
  $$ 4=\{\O, \{\O\}, \{\O,\{\O\}\},\{\O,
\{\O\}, \{\O,\{\O\}\} \} \}, {\kern 2pt} ... \ \ \ \ \ \ \ \ \ \ \
\ \
  $$

A Von Neumann set is Mirimanoff-first kind since it is isomorphic
with none of its parts. However this construction, associated with
the application of Morgan's laws to (\O), allowed the empty set to
be attributed an infinite descent of infinite descents and thus to
be classified as a member of the hypersets family (Bounias and
Bonaly, 1997b).

\medskip
\noindent\textbf{Proposition 3.4.} In a ordered pair
\{\{x\},\{x,y\}\}, the paired part \{x,y\} is unordered.

A classical acceptance (Schwartz, 1991) states that \{a,a\} =
\{a\}. This may introduce a confusion which can then be treated
from relation (5.1a) :
  $$
\rm (aa) = (xy) \ \Leftrightarrow \  a=x, \ a=y.
  \eqno(8.1 \rm b)
  $$

 This just means that for any x,y, a pair (a,a) is not
ordered. In effect, it becomes ordered upon rewriting: $\rm A =
(a,{\kern 2pt} (a,a))$. Comparing (aa) and A through relation
(8.1a), it  comes: $\rm (aa) = \{a, \{a,a\}\} \ \Leftrightarrow \
a=a$, and a=\{a,a\}. \quad \quad \quad \quad \quad   (Q.E.D.)

Thus: \{a,b\} = \{b,a\}, while $\rm (ab) \neq (ba)$ and $\rm
[\{a,a\} = a] \neq [ \{a, {\kern 2pt}(a)\}=(a,a)]$ where (a) is a
part, for which $\rm a \cup (a) \neq \{a\} \cup \{a\}$.
Consequently: (b,a) = (a,b) and (a,a) = !(a,a).

In a abstract set (in the sense of a set of any composition), one
can find parts with nonempty intersections as well as parts or
members with no common members. For instance, the following E and
F are respectively without order and at least partly ordered:
 $$
\rm E = \{a,{\kern 2pt} b, {\kern 2pt}\{c,d\},{\kern 2pt} G,
{\kern 2pt}X\} \ with \ a\neq b, \  c\neq d, \ G \neq X,
 \eqno(8.2)
 $$
 $$
\rm F = \{a, {\kern 2pt}b, {\kern 2pt}c, {\kern 2pt} \{a,d\},
{\kern 2pt}\{b,c,d\}, {\kern 2pt}\{d, {\kern 2pt} \{a,e,f\}\}\}.
 \eqno(8.3)
 $$

In particular cases, in a set like E or F, it may be that two
singletons, e.g. \{a\}, \{b\}, though different, have the same
weight with respect to a defined property. In such case one could
write that there exists some (m) such that $\rm \{a\}\equiv \{m\}$
and $\rm \{b\}\equiv \{m\}$, so that $\rm \{a, {\kern 2pt}d,
{\kern 2pt}G\} \cap \{c,{\kern 2pt} \{ b,c\}, {\kern 2pt}\}
=(a\equiv b)$.

\medskip
\noindent \textbf{Proposition 3.5.} An abstract set can be
provided with at least two kinds of orders: one with respect to
identification of a max and or a min, and one with respect to
ordered N-uples. These two order relations become equivalent upon
additional conditions on the nature of involved singletons.

\medskip
\noindent (i) A Mirimanof set of the (2.2) type and derived forms
can be provided a order. Let E = \{e, \{f,G\}\}. Then as seen
above, there exists some m such that $\rm \{e \} \equiv \{f\}
\equiv \{m\}$, and E is similar to $\rm E^\prime \equiv \{m,
{\kern 2pt}\{m,G\} \}$. Then, for any $\rm G \neq \O$, $\rm m
\subset \{m,G\}$ and thus m is a minimal member of \{m,G\}. Pose
$\rm \{m,G\}=M$, then $\rm E^\prime$ becomes $\rm E^{\prime
\prime}=\{m,M\}$ of which m is the minimal and M is the maximal of
the set. Then, one can rewrite $\rm \{m, \{m,G\} \}$ in the
alternative form $\rm \{m,{\kern 2pt} (m,G)\}$ and $\rm E =
\{e,{\kern 2pt} (f,G)\}$ since (f,G) is in some sort ordered.
These notations will be respected below. Note that $\rm
\angle(f,G) \Rightarrow \exists\{f,\{G\}\}$ while the reciprocal
is not necessarily true. The necessary condition is the following.

Suppose $\rm \{f,\{G\}\} \Rightarrow \exists\{f,\{G\}\}$. Then,
since $\rm \{f,\{G\}\}=\{\{G\},f\}$, one should be allowed to
write $\rm f\equiv G$ and $\rm G \equiv f$, that is, there is
neither minimal nor maximal in the considered part. Now, writing
(f,G) imposes as a necessary condition that there exists some m
and n such that $\rm (f,G) \equiv \{m,{\kern 1pt} \{m,n\}\} =
\{m,{\kern 1pt}\{n,m\}\}$. This can then be (but only
speculatively) turned into a \textit{virtual} (m,(m,n)). When
comparing two sets, parts ordered by this way will have to be
compared two by two: $\rm \{e,\{f,G\} \} =\{m,(n,Z)\} \Rightarrow
e = m$, and since $\rm m \neq M : f=n,{\kern 2pt} G=Z$ (otherwise
$\rm \{f,g\} = (n,z)$ could give alternatively $\rm f=z, {\kern
2pt} n=g$).

This drives the problem to the identification of orders in the set of parts
of a set, as compared with components of a simplex set (Table 1).

\medskip
\noindent (ii) A set can be ordered through rearrangement of its
exact members and singletons in a way permitted by the structure
of the set of parts of itself. In effect, existence of a set
axiomatically provides existence to the set of its parts
(Bourbaki, 1990a, p. 30) though an axiom of availability has been
shown to be required for disposal of the successive sets of parts
of sets of parts (Bounias, 2001).

Now, how to identify preordered pairs in a set? The set of parts
$\rm \mathcal{P}(E)$ of a set E provides the various ways members
of this set can be gathered in subsets, still not ordered.
Therefore, an analysis of the members of a set can involve a
rearrangements of the singletons contained in the set in a way
permitted by the arrangements allowed by $\rm \mathcal{P}(E)$,
including when a same singleton is present several times in the
set. Then, this may let emerge configurations that can be
identified with a structure of ordered N-uples.

\medskip
\noindent \textbf{Remark 3.3.} In N-uples as well as in the set of
parts, singletons gathered in any subpart are unordered. Call $\rm
(E^\prime)$ a set reordered by using $\rm \mathcal{P}(E)$. Then:
$\rm E^{\prime} {\kern 1pt}\cap {\kern 1pt} \mathcal{P}(E)$ $=
\mbox{\footnotesize{\textbf{Partition}}}(E)$.

\medskip
\noindent{\bf Application to simplicial structures}

Table 1 illustrates for the first four simplexes, the comparison
of the ordered pairs of the simplexes with the remaining part of
the set of parts, which is constituted of ($\rm 2^{N} - N - 1$)
unordered nonempty members for a simplex of N vertices.

\bigskip
\noindent \textbf{Table 1.} Sets and simplexes: N and D are the
numbers of vertices and the maximum dimension, respectively. $\rm
\mathcal{P}(E${\footnotesize{n}}) denotes the set of nonempty
parts of each {\footnotesize{n}} considered. Each class of subsets
of k members appears in $\rm C^{\kern 1pt k}_N$ forms. One of each
is used in a ordered n-uple, so that $\rm C^{\kern 1pt k}_N - 1$
are remaining. Finally, $\rm \sum_{{\kern 1pt} n=1 \rightarrow N}
(C^{\kern 1pt k}_n - 1) + N = 2^{\kern 1pt N}$.

\noindent
\newcommand{\PreserveBackslash}[1]{\let\temp=\\#1\let\\=\temp}
\let\PBS=\PreserveBackslash
\begin{longtable}
{|p{82pt}|p{20pt}|p{64pt}|p{28pt}|p{128pt}|p{38pt}|} a & a & a & a
& a & a  \kill \hline \medskip  \ \
\quad\quad\quad\quad\quad\quad\quad\quad
 \ \ Sets \ \ \  & \ \ \ \
\ \ Ver- \ ti- \ ces \ \ & \ \ \ \
\quad\quad\quad\quad\quad\quad\quad\quad \ \ \ \ \ \ \ \ \ \
Ordered
\par  pairing (N) &  \quad\quad\quad\quad\quad
Dim-ensio-nality & \quad\quad\quad\quad\quad\quad\quad\quad \ \ \
\ \ \ \ \ \ \ \ \ \ \ \ \ \ \ \ \ \ \ \ \ \ \ \ \ \ \ \

Remaining of $\rm \mathcal{P}(E{\footnotesize{n}})$
\par
 $(2^{\rm N} - {\rm N} - 1)$
\ \ \ \ \ \ \ \ \

&

\medskip

 ($2^{\rm N}-1)$
\quad \quad\quad \quad
 \\
\hline \ \ \ \quad\quad\quad\quad\quad\quad\quad\quad \ \ \ \ \ \
\ \ \ \ \quad\quad\quad\quad\quad\quad\quad\quad \ \ \ \ \ \ \ \ \
\
 \footnotesize{E1 = \{x1\}  \par
\quad\quad\quad\quad\quad\quad\quad\quad  \

  E2 = \{x1, x2\}
\par

\bigskip

\bigskip

 E3=\{x1,x2,x3\}
\par
\quad\quad\quad\quad\quad\quad\quad\quad
\quad\quad\quad\quad\quad\quad\quad\quad
\quad\quad\quad\quad\quad\quad\quad\quad

\medskip

 E4=\{x1,x2,x3,x4\} }

\quad\quad\quad\quad\quad\quad\quad\quad

&

\footnotesize{ \quad\quad\quad\quad\quad\quad\quad\quad
 N=1 \par
\quad\quad\quad\quad\quad\quad\quad\quad
  N=2 \par
\quad\quad\quad\quad\quad\quad\quad\quad
\quad\quad\quad\quad\quad\quad\quad\quad
\quad\quad\quad\quad\quad\quad\quad\quad

  N=3
  \par
\quad\quad\quad\quad\quad\quad\quad\quad
\quad\quad\quad\quad\quad\quad\quad\quad
\quad\quad\quad\quad\quad\quad\quad\quad

\medskip
  N=4 \par}

  &

\footnotesize{ \quad\quad\quad\quad\quad\quad\quad\quad
  \{x1\} \par
\quad\quad\quad\quad\quad\quad\quad\quad
  \{x1\},(x1,x2)  \par = (x1x2)
   \par
\quad\quad\quad\quad\quad\quad\quad\quad
   \{x1\},(x1,x2), \par (x1,x2,x3)
   \par = (x1x2x3)
\par

\medskip

 \{x1\},(x1,x2), \par
(x1,x2,x3), (x1,x2,x3,x4)
\par = (x1x2x3x4) }

&

\footnotesize{ \quad\quad\quad\quad\quad\quad\quad\quad
  D=0
 \par

\quad\quad\quad\quad\quad\quad\quad\quad
 D=1
 \par

\bigskip

\medskip

\medskip

 D=2
 \par

\bigskip

\bigskip

\medskip

 D=3 }

 &

\footnotesize{ \quad\quad\quad\quad\quad\quad\quad\quad

 none
\par

\bigskip

\{x2\}
\par

\bigskip

\bigskip

\{x2\},\{x3\},\{x1,x3\},(x2,x3)  \par

\bigskip

\bigskip

\medskip

\{x2\},\{x3\},\{x4\},
\par \{x1,x3\},\{x1,x4\},\{x2,x3\}, \par \{x2,x4\}, \{x3,x4\},
\{x1,x2,x4\}, \par \{x1,x3,x4\}, \{x2,x3,x4\}}

&

\footnotesize{

\hsize 5mm \ \ \   1
\par

\bigskip

 \ \ \ 3
\par

\bigskip

\bigskip \ \ \  7
\par

\bigskip

\bigskip

\medskip

\ \  15 }

\ \ \ \

\\ \hline
\end{longtable}

\bigskip

The remaining parts can in turn be arranged into appearing ordered
subparts. Hence for E3: $\rm \big\{ \{x2\},\{x2,x3\} \big\} =
(x1x3)$ and the \textit{"in} \textit{fine"} ordered $\rm \big\{\{
x3 \},\{x1,x3\} \big\} =(x3xl)$ since \{x1,x3\}=(x3,x1\}. For E4,
one gets $\rm \big\{ \{x2\},\{x2,x3\},\{x2,x3,x4\} \big\} =
(x2x3x4)$, $\rm \big\{ \{x3\},\{x3,x4\} \big\} = (x3x4)$, $\rm
\big\{ \{x1,x3\},\{x1,x3,x4\}\} = ((x1,x3){\kern 1pt}x4)$, and the
\textit{"in fine"} ordered: $\rm \big\{ \{x4\},\{x1,x4\} \big\} =
(x4x1)$, $\rm \big\{\{x2,x4\},\{x1,x2,x4\} \big\} = ((x2,x4){\kern
1pt}x1)$.

This further justifies that for a set like (E{\footnotesize{n}}),
the set of its ($2^{{\kern 1pt}\rm n}$) parts has dimension $\rm D
\leq D(E^{n})$).

{\it In fine}, a fully informative measure should provide a
picture of a space allowing its set component to be depicted in
terms of ordered N-uples ($\rm N = 0 \rightarrow n, {\kern 3pt}
n\subset \mathbf{N}$).

These considerations are intended to apply when a batch of observational
data and measurement are displayed in a scattered form on the laboratory
table of a scientist, and must then be reconstructed in a way most likely
representative of a previously unknown reality.

\subsubsection*{3.3.3. Identification of a "scanning"
measure on an abstract set}

\hspace*{\parindent} Several kinds of measure of a set, including
various forms of its diameter infer from section 2.1 above.

\medskip
\noindent \textbf{Proposition 3.6.} A set can be scanned by the
composition of a identity function with a difference function. Let
E = \{a,b,c,...\} a set having N members.

\medskip
\noindent(i) An identity function Id maps any members of E into
itself: $\rm \forall x \in E$, $\rm Id(x)=x$.

Thus, $\rm \angle (a, \ or \ b, \ or \ c, {\kern 2pt}...)$ this
provides one and only one response when applied to E. \\

\medskip
\noindent
(ii) A difference function is $f$ such that:  $\rm
\forall x \in E$, $f^{\rm n}(\rm x) \neq x$.

The exploration function is a self-map $\mathbf{M}$ of E:
$\mathbf{M}$ : $\rm E \mapsto E$, $\rm \mathbf{M} = Id \perp {\it
f}$:

 $$
\rm \angle {\kern 1pt} x \in E, \ \mathbf{M}(x)={\it f}(Id(x)), \
\forall n \ : \ \mathbf{M}^n(x) = {\it f}^n(Id(x)) \neq x.
 \eqno(9.1)
 $$

\medskip
\noindent \underline{Proof.} (a) Suppose $\rm \mathbf{M}=Id(x)$;
then, each trying maps a member of E to a fixed point and there is
no possible scanning of E. (b) Suppose one poses just $f
\rm(Id(x))\neq x$; then, given $f \rm(Id(x))\neq x$, say $f^{\kern
1pt 1}\rm(Id(x) = y$, since $\rm y \neq x$, then one may have
again $f^{\kern 1pt 2}\rm(Id(x)) = x$. Therefore, there can be a
loop without further scanning of E, with probability $\rm
(N-l)^1$. (c) Suppose one poses ${\rm M} = f$, such that $f\rm (x)
\neq x$. Then, since $f^{\kern 1pt 0}\rm (x) \neq x$, there can be
no start of the scanning process. If in contrast one poses as a
modification of the function {\it f} that $\rm \angle {\kern
1pt}x$, $f^{\kern 1pt 0} \rm (x) = x$, this again stops the
exploration process, since then $f^{\kern 1pt 0}\in \rm Id$. This
has been shown to provide a minimal indecidability case (Bounias,
2001).

The sequence of functions $\rm M^{n}(x) = {\it f}^{\kern 1pt n
}(Id(x)))\neq x$, $\rm \forall n$, is thus necessary and
sufficient to provide a measure of E which scans N-l members of E.
The sequence stops at the N$^{\rm th}$ iterate if in addition:
  $$
\rm {\it f}^{\kern 1pt n}(Id(x)) \neq \{{\it f}^{{\kern 1pt} i}
(Id(x)) \}_{\forall {\kern 1pt} i{\kern 1pt} \in {\kern 1pt}
[1,N]}.
  \eqno(9.2)
  $$

The described sequence represents an example of a path as
described above in more general terms.

Now, some preliminary kinds of diameters can be tentatively deduced for the
general case of a set E in which neither a complete structure nor a total
order can be seen.

\subsubsection*{3.3.4. Tentative evaluations of the size of sets
with ill-defined structure and order}

\hspace*{\parindent} Since in this case any two members of E are
of similar weight, regarding the definition of a diameter (4.3d),
it can be proposed the following definition.

\medskip \noindent
\textbf{Definition 3.1{\kern 0.5pt}a.} Given a non-ordered set E,
Id the identity self map of E, and {\it f} the difference self map
of E, a kind of diameter is given by the following relation:
  $$
\rm diam{\kern 1pt} {\it f}(E) \approx \{(x,y) \in E : \max {\it f
}^{\kern 1pt i}(Id(x)) \cap \max {\it f}^{\kern 1pt i}(Id(y) \}.
  \eqno(10.1)
  $$

This gives a (N-2) members parameter.

\medskip\noindent
\textbf{Subdefinition 3.1{\kern 1pt}b.} If E is well-ordered, i.e.
at least one, e.g. the lower boundary can be identified among
members of E, such as a singleton \{m\}, then an alternative form
can be written. Since in this case, the set E can be represented
by two members: $\rm E = \{m,Z\}$ with $\rm Z = \complement_{\kern
1pt E} (m)$ the complementary of m in E, relation (4.3d) results
in a measure $\rm M_{\kern 1pt H \Delta}$:
  $$
\rm M_{\kern 1pt H \Delta}(E) = \{ m = \min (E), {\kern 2pt}
\Delta (m,Z) \}
 \eqno(10.2)
  $$
with $\Delta$ the symmetric difference.

This gives a (N-1) member parameter which in turn provides a
derived kind of diameter $\rm diam_{\kern 1pt H \Delta}$ by
repeating the measure for two members m$^\prime$ and m$^{\prime
\prime}$ as on Figure 4. Let $\rm E^{\prime} = \{m^\prime,{\kern
2pt} \complement_{E}(m^\prime) \}$, and $\rm E^{\prime\prime}: \{
\complement_{E}(m^{\prime\prime})\}$, then (Figure 5):
  $$
\rm diam(E) \subseteq \{m^{\prime},m^{\prime\prime} \in E : \max
\Delta (E^\prime, E^{\prime\prime}) \}.
 \eqno(10.3)
  $$

\begin{figure}
\begin{center}
\includegraphics[scale=1.4]{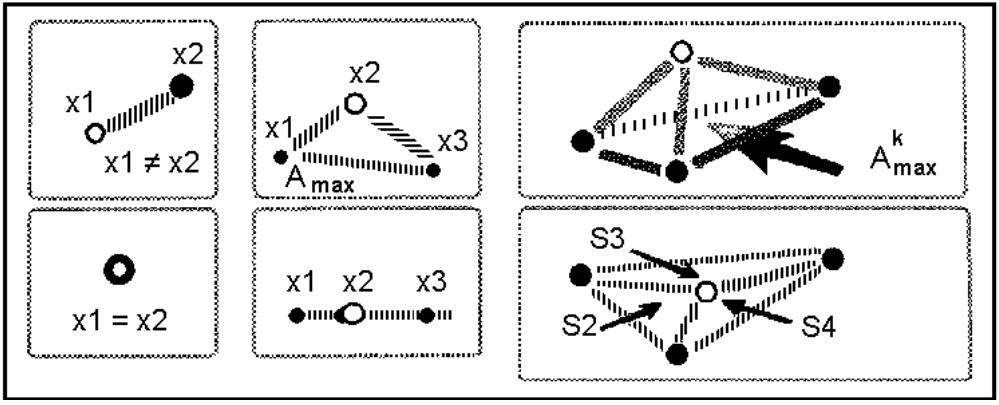}
\caption{\small{ The first three steps of the N-angular strict
inequality for the assessment of the dimensionality of a simplex.
In the lower right picture, the larger side standing for $\rm
A^{\kern 1pt k}_{max}$ is S1 such that in a 2-D space one has
exactly: $\rm S1=S2+S3+S4$.}} \label{5}
\end{center}
\end{figure}

\medskip\noindent
\textbf{Subdefinition 3.1c.} If both upper and lower boundaries
can be identified, i.e. the set E is totally ordered, then the
distance separating two segments A and B of E is:
  $$
\rm (A,B)\subset E, {\kern 2pt} dist_{E}(A,B) = \{ dist ( \inf
{\kern 1pt} A,{\kern 2pt} \inf{\kern 1pt} B) {\kern 2pt}\cap
{\kern 2pt } dist (sup {\kern 1pt } A,{\kern 2pt} sup {\kern 1pt}
B) \}
 \eqno(10.4)
 $$
where a set distance $\Delta$ is again provided by the symmetric
difference or by a n-D Borel measure.

A separating distance is a extrinsic form of the set-distance as an
intrinsic form.

A diameter is evaluated on E as the following limit:
 $$
\rm diam(E) = \{ \inf {\kern 1pt} A \rightarrow \inf {\kern 1pt}
E, {\kern 2pt} \sup  {\kern 1pt} A \rightarrow  \min {\kern 1pt}
E, \quad\quad\quad\quad\quad\quad\quad\quad\quad\quad \ \
 $$
 $$
\quad\quad\quad \rm \inf B \rightarrow \max E,  {\kern 1pt} \sup B
{\kern 1pt} \rightarrow \sup E  {\kern 2pt} | {\kern 2pt} \lim
{\kern 1pt} dist_{ {\kern 1pt}E}{\kern 1pt}(A,B) \}
 \eqno(10.5)
 $$

These preliminary approaches allow the measure of the size of
tessellating balls as well as that of tessellated spaces, with
reference to the calculation of their dimension through relations
(2.1)-(2.2) and (11.1)-(11.2) derived below. A diagonal-like part
of an abstract space can be identified with and logically derived
as, a diameter.

\medskip
\noindent\textbf{Remark 3.4.} If measure obtained each time from a
system, this means that no absolutely empty part is present as an
adjoined segment on the trajectory of the exploring path. Thus, no
space accessible to some sort of measure is strictly empty in a
both mathematical and physical sense, which supports the validity
of the quest of quantum mechanics for a structure of the void.

\subsubsection*{3.3.5. The dimension of an abstract space:
tessellatting with simplex k-faces}

\hspace*{\parindent} A major goal in physical exploration will be
to discern among detected objects which ones are equivalent with
abstract ordered N-uples within their embedding space.

A first of further coming problems is that in a space composed of
members identified with such abstract components, it may not be
found tessellatting balls all having identical diameter. Also a
ball with two members would have no such diameter as defined in
(10.1) or (10.4). Thus a measure should be used as a probe for the
evaluation of the coefficient of size ratio (p) needed for the
calculation of a dimension.

Some preliminary proposed solutions hold on the following
principles.\\

\noindent(i). A 3-object has dimension 2 iff given $\rm A^1_{max}$
its longer side fulfills the condition, for the triangular strict
inequality, where M denotes an appropriate measure:
 $$
\rm M(A^1_{max}) < M(A^1_2) + M(A^1_3).
 \eqno(11.1 \rm a)
 $$
Similarly, this condition can be extended to higher dimensions
(Figure 5):
 $$
\rm M(A^2_{max}) < M(A^2_2) + M(A^2_4)
 \eqno(11.1 \rm b)
 $$
Then, more generally for a space X being a N object:
 $$
\rm M(A^k_{max})< \mathop { \cup} \limits_{i = 1}^{N - 1} \{\
M(A^k_i) \}
 \eqno(11.1 \rm c)
 $$
with N = number of vertices, i.e. eventually of members in X, $\rm
k = (d-1) = N-2$, and A$\rm ^{k}_{max}$ the k-face with maximum
size in X. This supports a former proposition (Bounias, 2001).

\noindent \textbf{Remark 3.5.} One should note that, according to
relation (11.1c), for N = 2 (a "2-object"), X = \{x1, x2\} has
dimension 1 iff $\rm x1 < (x1 + x2)$, that is iff $\rm x1 < x2$
(Figure 5). This qualifies the lower state of an existing space
X$^{1}$.     \\

\noindent (ii). Let now space X be decomposed into the union of
balls represented by D-faces $\rm A^D$ proved to have dimension
$\rm Dim {\kern 2pt} (A^{D}) = D$ by relation (11.1c) and size
$\rm M(A^1)$ for a 1-face. Such a D-face is a D-simplex Sj whose
size, as a ball, is evaluated by $\rm M(A^1_{max})^D = Sj^{D}$.
Let $\mathcal{N}$ is the number of such balls that can be filled
in a space H, so that
 $$
\rm \mathop { \cup} \limits_{i = 1}^{\mathcal{N}} {\kern 1pt} \{
Sj^{D} \} \subseteq (H \approx L_{{\kern 1pt} max}^d)
 \eqno(11.2 \rm a)
 $$
with H being identified with a ball whose size would be evaluated
by $\rm L^{d}$, L is the size of a 1-face of H, and d the
dimension of H. Then, if $\rm \forall {\kern 2pt} Sj,{\kern 4pt}
Sj {\kern 1pt}\approx S{\kern 0.7pt}$o, the dimension of H is:

 $$
\rm Dim(H) \approx (D \cdot Log{\kern 2pt} S{\kern 0.7pt}o  +
Log{\kern 2pt} {\mathcal {N}}) {\kern 1pt}/ {\kern 2pt} Log {\kern
1pt} L^1_{{\kern 1pt} max}.
 \eqno(11.2 \rm b)
 $$

\medskip \noindent
\textbf{Remark.} Relation (11.2{\kern 0.5pt}b) stands for a kind
of interior measure in the Jordan's sense. In contrast, if one
poses that the reunion of balls covers the space H, then, Dim(H)
rather represents the capacity dimension, which remains an
evaluation of a fractal property.

\begin{figure}
\begin{center}
\includegraphics[scale=1.5]{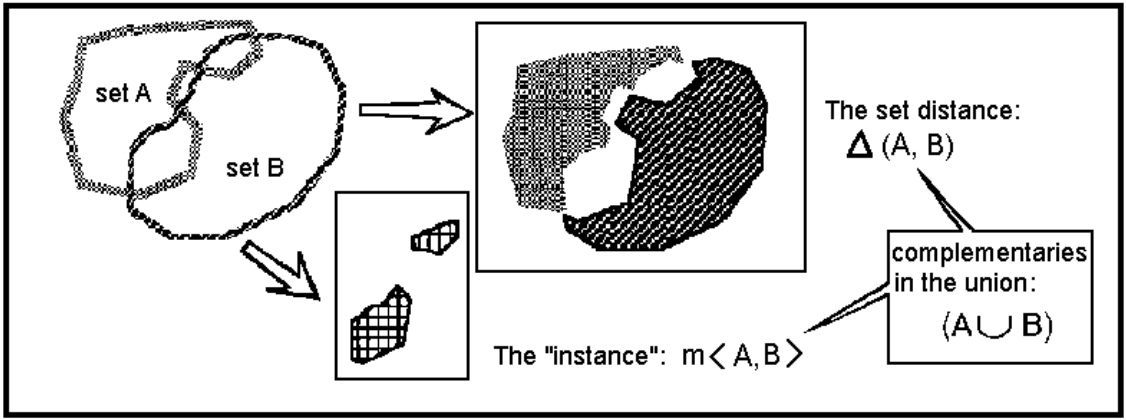}
\caption{\small{The symmetric difference ($\Delta$) between two or
more sets provides a kind of distance which is consistent with a
topology. This distance therefore stands for a topological
distance and provides topological spaces with a general form of
metrics. By this way, any topological space may be treated like a
(delta-) metric space. }} \label{6}
\end{center}
\end{figure}

\medskip \noindent
\textbf{Definition 3.2{\kern 0.5pt}b.} A n-(simplex-ball) is the
topological unit ball circumscribed to a n-simplex. \\

\noindent (iii). Extension to a ill-defined space E.

The problem consists in identifying first a 1-face component of E
from k-faces ($\rm k>l$), which implies to identify ordered
N-uples. Then, components of E, whatever their nature, should be
analogically decomposed into appropriate simplexes. The number of
these simplexes will then be evaluated in E, and relations (11.2)
will be finally applied.

\medskip \noindent \textbf{Remark.} There is no other condition on
simplexes Sj than the need that one face is maximal: this is the
face on which the others will be projected, so that the
generalized inequality will be applied (Figure 6).

Depending on $\mathcal{N}$, noninteger d(H) can be obtained for a
fractal or a fractal-like H. Some applications of the above
protocol will be illustrated in further communications.

\medskip \noindent
\textbf{Lemma 3.2.} A singleton or member (a) is putatively
available in the form of a self-similar ordered N-uple: $\rm \{a\}
= \{a \cap a \cap {\kern 2pt}... {\kern 2pt} \cap a\}_h \neq \{
(a,(a),{\kern 1pt}...\}$.

In effect, the set theory axiom of the reunion states that $\rm A
\cup A = A$ as well as the axiom of the intersection states that
$\rm A \cap A = A$.

\medskip
\noindent \textbf{Corollary 3.2.} In the analysis of a abstract
space $\rm H = L^{x}$, of which dimension x is unknown, the
identification of which members can be identified with N-uples
supposedly coming from a putative Cartesian product of members of
H, e.g.: $\rm G \subset H$: $\rm \angle {\kern 1pt} a \in G$, $\rm
(aaa...a)_h \in G^{h}$, is allowed by an anticipatory process.

\medskip \noindent
\underline{Preliminary proof.} At time (t) of the analysis of the
formal system involved, there is no recurrent function that can
"imagine" the existence of an abstract component not existing in
the original data and parameters, and not directly inferring from
a computation of these data and parameters. Devising $\rm
(aaa...a)_h \in G^{h}$, implies making a mental image at $\rm
(t+te)$ and further confronting the behavior of the system with
"anticipatively recurrent" images succeeding to those at t, that
is computed from (t, $\rm t+e$) conditions within (t+1, t+2, ...,
up to $\rm t+e-1$).

Let $\rm E = \{a,{\kern 2pt} \{b,{\kern 1pt}\{,B,C\}\}\}$; then
one anticipates on the putative existence of a unordered pair $\rm
\{a,a\} = a$ in a former writing of E.

Therefore, $\rm (a) = \big\{ \{a,{\kern 1pt}\{a,a\}\}{\kern 1pt} |
{\kern 1pt} \{a,a\} = \{a\} \big\}$. The presence of singletons
can be identified with a putative former reduction of ordered
parts owning the same nucleus (a). A nucleus thus appears as a
ordered form of a singleton, that is the only case where a ordered
form is identical with a unordered one.

\medskip \noindent
\textbf{Application 3.2.} Given ($\rm E^{{\kern 1pt}e}_{{\kern
0.5pt}n}$) composed of N-uples denoted ($\rm \Pi_{{\kern 0.5pt
i}}^{{\kern 0.5pt} e}$) in their ordered state, and $\rm K ^e_i$,
$\rm K^e_j$, ... , in their unordered acceptation: an approach of
the identification of the diagonal, and thus of a further measure
of the respective diameters of ($\rm E^{{\kern 0.5pt}e}_{{\kern
0.5pt}n}$) and $\rm \Pi_{{\kern 0.5pt i}}^{{\kern 0.5pt} e}$) is
given by the following propositions:
 $$
\rm diag{\kern 1pt} (E^{{\kern 0.5pt}e}_{{\kern 0.5pt}n})
\subseteq \min \cup {\kern 1pt}\{ (K^{{\kern 0.5pt}e}_{{\kern
0.5pt}i},{\kern 2pt} K^{{\kern 0.5pt} e}_{{\kern 0.5pt}j}) \in
E^{{\kern 0.5pt}e}_{{\kern 0.5pt}n} : \Delta(K^{{\kern
0.5pt}e}_{{\kern 0.5pt}i},{\kern 2pt} K^{{\kern 0.5pt} e}_{{\kern
0.5pt}j})_{{\kern 1pt}i\neq j} \}
 \eqno(11.3 {\kern 0.5pt} \rm a)
 $$
 $$
\rm diag (\Pi_{{\kern 0.5pt i}}^{{\kern 0.5pt} e})  \subseteq \max
\cup {\kern 1pt}\{ (K^{{\kern 0.5pt}e}_{{\kern 0.5pt}i},{\kern
2pt} K^{{\kern 0.5pt} e}_{{\kern 0.5pt}j}) \in  E^{{\kern
0.5pt}e}_{{\kern 0.5pt}n} : \Delta(K^{{\kern 0.5pt}e}_{{\kern
0.5pt}i},{\kern 2pt} K^{{\kern 0.5pt} e}_{{\kern 0.5pt}j})_{{\kern
1pt}i\neq j} \}
 \eqno(11.3 {\kern 0.5pt} \rm b)
 $$

An example is given in Appendix as a tentative application case.
Regarding just a Cartesian product, the set of parts with empty or
minimal intersections stands for the diagonal and the diagonal of
a polygon stands for a topological diameter.

\section*{4. Physical inferences}

\subsection*{4.1. Defining a probationary space}

\hspace*{\parindent} A probationary space (Bounias, 2001) is
defined as a space fulfilling exactly the conditions required for
a property to hold, in terms of:  \\

(i) identification of set components;

(ii) identification of combinations of rules

(iii) identification of the reasoning system.   \\

All these components are necessary to provide the whole system
with decidability (Bounias, 2001). Since lack of mathematical
decidability inevitably flaws also any physical model derived from
a mathematical background, our aim is to access as close as
possible to these imposed conditions in a description of both a
possible and a knowable universe, in order to refine in
consequence some current physical postulates.

The above considerations have raised a set of conditions needed for some
knowledge to be gettable about a previously unknown or ill-known space, upon
pathwise exploration from a perceiving system A to a target space B within
an embedding space B.

In the absence of preliminary postulate about existence of
so-called "matter" and related concepts, it has been demonstrated
that the existence of the empty set is a necessary and sufficient
condition for the existence of abstract mathematical spaces
(W$^{\rm n}$) endowed with topological dimensions (n) as great as
needed (Bounias and Bonaly, 1997). Hence, the empty set appears as
a set without members though containing empty parts. The reasoning
intermediately proved that the empty set owns the properties of a
nonwellfounded set and exhibit (i) self-similarity at all scales
and (ii) nowhere derivability, that is: two characteristics of
fractal structures. These properties have been shown to clear up
some antinomies remained unaddressed about the empty set
properties.

These findings will now be shown to provide additional features of physical
interest. In first, the empty set can provide at least an intellectual
support to existence of some sort of space.

\subsection*{4.2. The founding element}

\hspace*{\parindent} It is generally assumed in textbooks of
mathematics (see for instance Schwartz, 1991, p. 24), that some
set does exist. This strong postulate has been reduced to a weaker
form reduced to the axiom of the existence of the empty set
(Bounias and Bonaly, 1997). It has been shown that providing the
empty set (\O) with ($\in, \subset$) as the combination rules
(that is also with the property of complementarity
($\complement$)) results in the definition of a magma allowing a
consistent application of the first Morgan's law without violating
the axiom of foundation iff the empty set is seen as a hyperset,
that is a nonwellfounded set (Aczel, 1987, Barwise, 1991). A
further support of this conclusion emerges from the fact that
several paradoxes or inconsistencies about the empty set
properties are solved (Bounias and Bonaly, 1997). These
preliminaries now drive us to the formulation of the following
theorem which will be established using several Lemmas.

\subsection*{4.3. The founding lattice}

\textbf{Theorem 4.1.} The magma $\O^{\O}=\{ \O,{\kern 1pt}
\complement \}$ constructed with the empty hyperset and the axiom
of availability is a fractal lattice.

\medskip \noindent \textbf{Remark.} The notation $\O^{\O}$ has been
preferred to -say- ($\rm E_{\O}$) or others. In effect, in
Bourbaki notation, $\rm E^{F}$ means the space of functions of set
F in set E. Therefore, writing ($\O^{\O}$) denotes that the magma
reflects the set of all self-mappings of (\O), which emphasizes
the forthcoming results.

\medskip \noindent
\textbf{Lemma 4.1.} The space constructed with the empty set cells
of $\rm E${\scriptsize {\O}} is a Boolean lattice.

\medskip \noindent
\underline{ Proof.} (i) Let $\rm \cup(\O) = S$ denote a simple
partition of (\O). Suppose that there exists an object
($\varepsilon$) included in a part of S, then necessarily
$(\varepsilon) = \O$ and its belongs to the partition.

(ii) Let $\rm P = \{\O, \O\}$ denote a part bounded by {\kern
0.2pt} $\rm sup{\kern 2pt} P = S$ {\kern 0.2pt} and {\kern 1pt}
$\rm inf {\kern 2pt}P = \{\O \}$. The combination rules $\cup$ and
$\cap$ provided with commutativity, associativity and absorption
are holding. In effect: $\O \cup \O = \O$, $\O \cap \O = \O$ and
thus necessarily $\O \cup (\O \cap \O) = \O$, $\O \cap (\O \cup
\O) = \O$. Thus, space $\rm \{ (P(\O), (\cup, \cap)\}$ is a
lattice.

The null member is {\O} and the universal member is $2^{{\kern
1pt}\O}$ that should be denoted by $\aleph_{{\kern 1.3pt}\O}$.
Since in addition, by founding property $\complement_{{\kern
1pt}\O}(\O)=\O$, and the space of (\O) is distributive, then $\rm
S(\O)$ is a Boolean lattice. \qquad \qquad \qquad \qquad  \quad
\quad    (Q.E.D.)

\medskip \noindent
\textbf{Lemma 4.2.} S(\O) is provided with a topology of discrete
space.

\medskip \noindent
\underline{Proof.}{\kern 1pt} (i) The lattice S(\O) owns a
topology. In effect, it is stable upon union and finite
intersection, and its contains (\O). \\

\noindent (ii){\kern 1pt} Let S(\O) denote a set of closed units.
Two units $\O_1$, $\O_2$ separated by a unit $\O_3$ compose a part
$\{\O1,{\kern 1pt} \O_2, {\kern 1pt} \O_3 \}$. Then, owing to the
fact that the complementary of a closed is a open: $\complement _
{{\kern 1pt} \{\O_1,{\kern 1pt} \O_2,{\kern 1pt} \O3 \}} {\kern
1pt} \{\O_1,{\kern 1pt} \O_3 \} = \O_2$, and $\O_2$ is open. Thus,
by recurrence, $\{\O1,{\kern 1pt} \O_3 \}$ are surrounded by open
]\O[ and in parts of these open, there exists distinct
neigborhoods for ($\O_1$) and ($\O_3$). The space $\rm S(\O)$ is
therefore Hausdorff separated. Units (\O) formed with parts thus
constitute a topology ($\mathcal T_{\O}$) of discrete space.
Indeed, it also contains the discrete topology $(\O^{\O}, {\kern
1pt}(\O))$, which is the coarse one and is of much less
mathematical interest.

\medskip \noindent
\textbf{Lemma 4.3.} The magma of empty hyperset is endowed with
self-similar ratios.

The Von Neumann notation associated with the axiom of
availability, applying on $(\O)$, provide existence of sets $\rm
(N^{\O})$ and $\rm (Q^{\O})$ equipotent to the natural and the
rational numbers (Bounias and Bonaly, 1997). Sets Q and N can thus
be used for the purpose of a proof. Consider a Cartesian product
$\rm E n \times E n$ of a section of $\rm (Q^{\O})$ of n integers.
The amplitude of the available intervals range from 0 to n, with
two particular cases: interval [0,{\kern 1pt}1] and any of the
minimal intervals $\rm [1/n-l, {\kern 2pt} 1/n]$. Consider now the
open section ]0,{\kern 1pt}1[: it is an empty interval, noted
$\O_1$. Similarly, note $\rm \O_{min} = ]0, {\kern 2pt}
l/n(n-l)[$. Since interval $\rm [0,{\kern 2pt} l/n(n-l)]$ is
contained in [0,{\kern 1pt}1], it follows that $\rm
\O_{min}\subset \O_1$. Since empty sets constitute the founding
cells of the lattice $\rm S(\O)$ proved in Lemma 4.1, the lattice
is tessellatted with cells (or balls) with homothetic-like ratios
of at least $\rm r = n (n-l)$. The absence of unfilled areas will
be further supported in Part 2 of this study by introduction of
the "set with no parts".

\medskip \noindent
\textbf{Definition 4.1.} Such a lattice of tessellation balls will
be called a "tessellattice".

\medskip \noindent
\textbf{Lemma 3.3.4.} The magma of empty hyperset is a fractal
tessellattice.

\medskip \noindent
\underline{Proof.} \\

\noindent (i) From relations (2.3) one can write $(\O)\cup (\O) =
(\O,\O) = (\O)$.

\noindent (ii) It is straightforward that $(\O) \cap (\O) = (\O)$.

\noindent (iii) Last, the magma $(\O^{\O}) = \{\O, \complement\}$
represents the generator of the final structure, since (\O) acts
as the "initiator polygon", and complementarity as the rule of
construction. These three properties stand for the major features
which characterize a fractal object (James and James, 1992).

Finally, the axiom of the existence of the empty set, added with
the axiom of availability in turn provide existence to a lattice
$\rm S(\O)$ that constitutes a discrete fractal Hausdorff space,
and the proof is complete.

\subsection*{4.4. Existence and nature of spacetime}

\textbf{Lemma 4.5.} A lattice of empty sets can provide existence
to a at least physical-like space.

\medskip \noindent
\underline{Proof.} Let {\O}  denote the empty set as a case of the
whole structure, and \{\O\} denotes some of its parts. It has been
shown that the set of parts of {\O} contains parts equipotent to
sets of integers, of rational and of real numbers, and owns the
power of continuum (Bounias and Bonaly, 1994,1997). Then, looking
at the inferring spaces ($\rm W^{n}$), ($\rm W^{m}$), ... thus
formed, it has been proved (Bounias and Bonaly, 1994) that the
intersections of such spaces having nonequal dimensions give raise
to spaces containing all their accumulation points and thus
forming closed sets. Hence:
 $$
\rm \{(W^{{\kern 1pt}n}) \cap (W^{{\kern 1pt}m}) \}_{m > n} =
(\Theta^{{\kern 1pt}n}) \ is \  closed \ space.
 \eqno(12)
 $$

These spaces provide collections of discrete manifolds whose
interior is endowed with the power of continuum. Consider a
particular case ($\Theta^{{\kern 1pt}4}$) and the set of its parts
$\mathcal {P}(\Theta^{{\kern 1pt}\rm n})$; then any of
intersections of subspaces $(\rm E^{{\kern 1pt}d})_{{\kern
1pt}d<4}$ provides a d-space in which the Jordan-Veblen theorem
allows closed members to get the status of both observable objects
and perceiving objects (Bounias and Bonaly, 1997b). This stands
for observability, which is a condition for a space to be in some
sort observable, that is physical-like (Bonaly's conjecture,
1992).

Finally, in any ($\Theta^{{\kern 1pt}4}$) space, the ordered
sequences of closed intersections $\{ (\rm E^{{\kern
1pt}d})_{{\kern 1pt}d<4}\}$, with respect to mappings of members
of  $\{(\rm E^{{\kern 1pt}d})_{{\kern 1pt}d<4} \}_i$ into  $\{(\rm
E^{{\kern 1pt}d})_{{\kern 1pt}d<4} \}_j$, provides an orientation
accounting for the physical arrow of time (Bounias, 2000a), in
turn embedding an irreversible arrow of biological time (Bounias,
2000b). Thus the following proposition:

\medskip \noindent
\textbf{Proposition 4.1.} A manifold of potential physical
universes is provided by the ($\Theta^{\kern 1pt \rm n}$) category
of closed spaces.

Our spacetime is one of the mathematically optimum ones, together
with the alternative series of $\rm \{(W^{3})\cap (W^{m})
\}_{in>3}$. Higher spacetimes ($\rm \Theta^{{\kern 1pt}n})_{{\kern
1pt} n}>3$ could exist as well.

Now it will be briefly examined some new structures which can pertain to a
topological space as described above, and deserves specific attention.

\section*{5. Towards "fuzzy dimension" and "beaver spaces"}

\subsection*{5.1. Introduction}

\medskip
\hspace*{\parindent} The exploration of nature raises the
existence of strange objects, such as living organisms, whose
anatomy suggests that mathematical objects having adjoined parts
with each having different dimensions could exist. This will then
introduce the idea that the dimensions of some objects may even
not be completely established. The existence of such strange
objects implies that appropriate tools should be prepared for
their eventual study. This is sketched here and will be matter of
further developments.

\medskip
\subsection*{5.2. From hairy spaces to "beaver spaces"}

\hspace*{\parindent}
 A "hairy space" is a $\rm (n>3)$-ball having 1-D
lines (also called "hair" of "grass") planted on it (Berger,
1990). It is interesting that for such spaces ($\rm n>3$ only),
the volume and the area do not change with the insertion of
1-spaces on them.

Let a simplex $\rm S_{n}^{n}$: the last segment allowing set $\rm
E_{{\kern 1pt}n+1}$ to be completed from $\rm E_{n}$ of $\rm
S^{{\kern 1pt} n-1}$ is $(\rm x_{{\kern 1pt} n},{\kern
1pt}x_{{\kern 1pt}n+})$. Consider the simplex $\rm F^{{\kern
1pt}n-1}$ such that one of its facets $\rm A^{n-2}$ has its last
segment $\rm (y_{{\kern 1pt}n-1},{\kern 1pt}y_{{\kern 1pt}n-2})
\equiv (x_{{\kern 1pt } n}, {\kern 1pt} x_{{\kern 1pt} n+1 })$.
Repeat this operation for descending values: one gets a space
having n- and $\rm (n-l)$-adjoined parts with their intersections
having lower dimension: $\rm Dim \{S^{{\kern 1pt} n} \cap
S^{{\kern 1pt}n-1} \} \leq (n-2)$. With respect to a beaver,
having a spherical body, with a flat tail surrounded with hair,
these spaces will be denoted as "beaver spaces". Other
possibilities include not ordered appositions of parts with
various dimensions.

The existence of Beaver spaces implies some specific adjustment of
the methods used for their scanning. In this respect, it should be
recalled that a new mode of assessment of coordinates has been
formerly proposed (Bounias and Bonaly, 1996): it consists in
studying the intersection of the unknown space with a probe
composed with an ordered sequence of topological balls of
decreasing dimensions, down to a point ($\rm D=0$). This process
has been shown to wear the advantage of being able to define
coordinates even in a fractal space. This may be of particular
interest with components provided by and embedded in the lattice
$\rm S(\O)$.

\subsection*{5.3. "Fuzzy dimensioned" spaces}

\textbf{Proposition 5.1.} There exist spaces with fuzzy
dimensions.

\medskip \noindent
\underline{Proof.} Consider a simplex $\rm S^{{\kern 1pt}
n}_{{\kern 1pt}n} = \{(E_{{\kern 1pt} n+}), {\kern 1pt}(\perp n)
\}$,  and its two characteristic structures, namely $\rm
\mathbf{L}^{1}_{{\kern 1pt}n} = \sum_{{\kern 1pt} i=1 \rightarrow
n} \big( dist(x_{{\kern 1pt} i-1},x_i) \big)$ {\kern 1pt} and $\rm
\yen_{{\kern 1pt}n+1}^1=\sum_{{\kern 1pt} i=1 \rightarrow n} \big(
dist(x_{{\kern 1pt}i+1},x_i)\big)$. Then, a condition for the
assessment of simplicial dimensionality is given by the structure
defining a simplicial space (Bounias, 2001) in consistency with a
$\rm (1-D)$-probe for a $\rm (D>l)$-space:
 $$
\rm (\perp n) = ( \yen_{n+1}^1 > k(n,d) \cdot \mathbf{L}_{n}^1
\Leftrightarrow  d > n).
 \eqno(13)
 $$

Let the last segment of the simplex $(\rm x_{n},{\kern
1pt}x_{n+1})$ be such that, in consistency with Zadeh and Kacprzyk
(1992), one has: $\rm dist (\rm x_{n},{\kern 1pt}x_{n+1})\in
[0,1]$. Then, the expression of $\rm \yen^1_{n+1}$ reaches a value
fuzzily situated between the assessment of $\rm D = n-1$ and $\rm
D=n$ at least for $\rm n < 4$. Thus, the simplicial set $\rm
E_{n+1}$ is of the fuzzy type and the simplex is a fuzzy simplex.

Such a space will therefore be said as having a "fuzzy dimension".
This provides a particular extension of the concept of fuzzy set
into that of "fuzzy space of magma", since the set is unchanged
while it is the rule that provides the magma with a fuzzy
structure.

These problems will be the matter of further developments, since
they belong to the kind of situations that could really be
encountered during the exploration of universe, and as pointed by
Klir and Wierman (1999): "Knowledge about the outcome of an
uncertain event gives the possessor an advantage".

\section*{6. Discussion}

\subsection*{6.1. Epistemological assessment}

\hspace*{\parindent} It has been stressed in the introduction that
computable systems can in principle not be self-evaluated, due to
incompleteness and indecidability theorems. It will be shown here
on a simple example why this does not apply to the self-evaluation
of our universe structure by one of its components, when this
component is a living subpart provided with conscious perception
functions, that is a brain.

\medskip
\noindent \textbf{Theorem 6.1.} A world containing subparts
endowed with consciously perceptive brains is a self-evaluable
system.

The proof will be founded on two main Lemmas.

\medskip \noindent
\textbf{Lemma 6.1.} There exists a minimal indecidable system
(Bounias, 2001).

\medskip \noindent
\underline{Proof.} The former theorem of Godel provided the most
sophisticated proof, while the systems devised by Chaitin
described the most giant examples of known indecidable systems. In
contrast, a recurrent reasoning would state that there should
exist most simpler cases. Consider the set $\rm E=\{(A)\}$ where
(A) is a part, eventually composed of nuclei or singletons. Try an
exploration of E by itself: without prior knowledge of the inside
structure of E, the exploration function $\rm {\it f}^{{\kern
1pt}n}(Id(x))$ described in relations (2.3.1--2.3.5). Applying
this function to E, one gets $\rm Id(A)=A$ and $\rm {\it
f}^{{\kern 1pt}n}(A)\neq A$ does not exists from the definitions
(this applies to $\rm A=\{a\}$ a singleton, as well). Hence, the
system returns no result: not a zero result, but litterally no
answer. Now, let A be composed of singletons, the same procedure
must be applied to each detected singleton, with the same failure.
Finally, let A be ($\O${\scriptsize{\O}}): for any singleton
\{\O\}, one will get $\rm Id(\O) = \O$ and $\rm {\it f}^{{\kern
1pt} n}(\O)$ again returns no response. Hence, the lattice S(0) is
not self-evaluable as it stands at this stage. Since \{\O\} is the
minimum of any nonempty set, the system stands for a minimum of an
indecidable case of the classical set theory (though a new
component will be considered in Part 2).

\medskip
\noindent \textbf{Lemma 6.2}. A consciously perceptive biological
brain is endowed with anticipatory properties (Bounias and Bonaly,
2001).

The function of conscious perception has been shown to infer from the same
conditions as those allowing a physical universe to exist from abstract
mathematical spaces: A path connecting a Jordan's point of an outside closed
B to the inside of another closed A is prolonged in a biological brain into
a sequence of neuronal configurations which converges to fixed points. These
fixed points stand for mental images (Bounias and Bonaly, 1996; Bounias,
2000). The sequence of mental images owns fractal properties which can
provide additional help to the construction by the brain of mental images of
an expected future state. These mental images will in turn be used as a
guide for the adjustment of further actions to the expected goal (Bounias,
2000), which is also a way by which the organism returns molecular
information to the brain, making unconscious (autonomous) mental images
which are used in turn for the control of the homeostasy of the organism.

Now, consider a space $\rm F=\{(A),(B)\}$ where (A) is a living
system endowed with anticipatory properties. Then, A is able to
analyze some of its components (a$_1$, a$_2$, ...) through a way
similar to the exploration function. By an anticipatory process,
it is able to construct the power set of parts $\rm {\mathcal
{P}}^{{\kern 1pt}n}$ of at least a part of (A) or (B).
Consequently, there will always be a step (n) in which $\rm
{\mathcal {P}}^{{\kern 1pt}n}$, i.e. the set of parts of the set
of parts of ... (n times) the set of parts of part of (A) or (B)
will have a cardinality higher than (A) or (B), since (A) and (B)
are finite and $\rm {\mathcal {P}}^{{\kern 1pt}n}$ is infinitely
denumerable. Then (A) will be able to construct a surjective map
of $\rm {\mathcal {P}}^{{\kern 1pt}n}$ on either (A) or (B). This
completes the proof.

\subsection*{6.2. About the assessment of probationary spaces}

\hspace*{\parindent} There remain enormous gaps in the present-day
knowledge of what universe could really be. For instance, current
cosmological theories remain contradictive with astronomical
observation (Mitchell, 1995; Bucher and Spergel, 1999; Krauss,
1999 and many other articles). The inconsistency of Lorentz
covariance used by Einstein, Minkowski, Mach, Poincar\'{e},
Maxwell etc. with Lorentz invariance used by Dirac, Wigner,
Feynman, Yang, etc. remains unsolved (Arunasalam, 1997). Quantum
mechanics still failing to account for macroscopic phenomena,
could be -and has been -interpreted through classical physics
(Wesley, 1995), while flaws have been found in so-called "decisive
experimental evidence" or classical observations about as
fundamental parameters as the Bell inequalities (Wesley, 1994),
the velocity of light (Driscoll, 1997), the red shift meaning
(Meno, 1998), and others.

More, whether space is independent from matter or matter is deformation of
space remains questioned (Krasnoholovets, 1997; Kubel, 1997; Rothwarf,
1998). These discrepancies essentially come from the fact that probationary
spaces supporting a number of explicit or implicit assessments have not been
clearly identified.

At cosmological scales, the relativity theory places referentials in a
undefined space, with undefined gauges nor substrate for the transfer of
information and the support of interactions. That matter exists and is
spread into this undefined medium is just implicitly admitted without
justifications. Here, distances are postulated without reference to objects.

At quantum scales, a probability that objects are present in a certain
volume is calculated. But again, nothing is assessed about what are these
objects, and what is their embedding medium in which such "volumes" can be
found. Furthermore, whether these objects are of a nature similar or
different to the nature of their embedding medium has not been addressed. In
this case, objects are postulated without reference to distances.

About quantum levels, justifications have been mathematically produced in
order to cope with some unexplainable observations, but this does not
constitute a proof, per se, since the proof is not independent from the
result to be supported. Last, neither the Big Bang energy source nor others
have been justified, which led to the theory of an energy of the void: but
then, this precludes existence of a true "void". These remarks support the
need for finding ultrafilter properties which would be provided to any
object and distance from microscopic to cosmic scales in our universe,
assuming that it is not composed of separate component with discontinuity
nor break of arcwise connectivity.

The next part of this work will provide some logical answers to such
problems and derive physical properties of an infering spacetime, with
particular reference to the derivation of cosmic scale features from
submicroscopic characteristics.

\subsection*{References}

Aczel, P., 1987. {\it Lectures on Non-Well-Founded Sets. CSLI
Lecture-notes 9}, Stanford, USA. \\

\noindent Arkani-Hamed, N., Dimopoulos, S., Dvali, G., 2000. "Les
dimensions cach\'{e}es de 1'univers", {\it Pour la Science}
(Scientific American: French edition), Vol. 276, pp. 56-65. \\

\noindent Arunasalam V., 1997. "Einstein and Minkowski versus
Dirac and Wigner: covariance versus invariance", {\it  Physics
Essays}, Vol. 10, No. 3, pp. 528-32. \\

\noindent Avinash, K., Rvachev, V.L., 2000. "Non-Archimedean
algebra: applications to cosmology and gravitation", {\it
Foundations of physics}, Vol. 30, No. 1, pp. 139-52. \\

\noindent Banchoff, T., 1996. {\it The fourth dimension}.
Scientific American Books Inc., French edition, Pour La
Science-Belin, Paris, 69-80. \\

\noindent Barwise, I, Moss, J. (1991). "Hypersets", {\it Math.
Intelligencer}, Vol. 13, No. 4, pp. 31-41. \\

\noindent Bonaly, A., 1992. Personal communication to M. Bounias.
\\

\noindent Bonaly, A., Bounias, M., 1995. "The trace of time in
Poincar\'{e} sections of topological spaces", {\it  Physics
Essays}, Vol. 8, No. 2, pp. 236-44. \\

\noindent Borel, E., 1912. "Les ensembles de mesure nulle", in
{\it Oeuvres compl\`{e}tes}. Editions du CNRS, Paris, Vol. 3.  \\

\noindent Bourbaki, N., 1990a. {\it Th\'{e}orie des ensembles},
Chap. 1-4. Masson, Paris, p. 352. \\

\noindent Bourbaki, N., 1990b. {\it Topologie G\'{e}n\'{e}rate},
Chap. 1-4. Masson, Paris, p. 376. \\

\noindent Bounias, M., Bonaly, A., 1994. "On mathematical links
between physical existence, observability and information: towards
a "theorem of something", {\it  J. Ultra Scientist of Physical
Sciences}, Vol. 6, No. 2, pp. 251-59.  \\

\noindent Bounias, M., Bonaly, A., 1996. "On metrics and scaling :
physical coordinates in topological spaces", {\it Indian Journal
of Theoretical Physics}, Vol. 44, No. 4, pp. 303-21. \\

\noindent Bounias, M., 1997. "Definition and some properties of
set-differences, instances and their momentum, in the search for
probationary spaces", {\it J. Ultra Scientist of Physical
Sciences}, Vol. 9, No. 2, pp. 139-45. \\

\noindent Bounias, M., Bonaly, A., 1997a. "The topology of
perceptive functions as a corollary of the theorem of existence in
closed spaces",  {\it BioSystems}, Vol. 42, pp. 191-205.  \\

\noindent Bounias, M., Bonaly, A., 1997b. "Some theorems on the
empty set as necessary and sufficient for the primary topological
axioms of physical existence", {\it Physics Essays}, Vol. 10, No.
4, pp. 633-43.  \\

\noindent Bounias, M., 2000a. "The theory of something: a theorem
supporting the conditions for existence of a physical universe,
from the empty set to the biological self", in (Daniel M. Dubois,
Ed.): CASYS'99 Int. Math. Conf., {\it Int. J. Comput. Anticipatory
Systems}, Vol. 5, pp. 11-24.  \\

\noindent Bounias, 2000b. A theorem proving the irreversibility of
the biological arrow of time, based on fixed points in the brain
as a compact, delta-complete topological space", in (Daniel M.
Dubois, Ed.) CASYS'99 Int. Math. Conf., {\it American Institute of
Physics CP}, Vol. 517, pp. 233-43.  \\

\noindent Bounias, M., 2001. "Indecidability and Incompleteness In
Formal Axiomatics as Questioned by Anticipatory Processes", in
(Daniel M. Dubois, (Ed.) CASYS'2000 Int. Math. Conf., {\it Int. J.
Comput. Anticipatory Systems}, Vol. 8, pp. 259-74.  \\

\noindent Bounias, M., Bonaly, A., 2001. "A formal link of
anticipatory mental imaging with fractal features of biological
time", {\it Amer. Inst. Phys. CP}, Vol. 573, pp. 422-36.  \\

\noindent Bucher, M., Spergel, D., 1999. "L'inflation de
1'univers", {\it Pour la Science} ({\it Scientific American},
French ed.), Vol. 257, pp. 50-7.  \\

\noindent Chaitin, G.J., 1998. {\it The Limits of Mathematics},
Springer-Verlag (Singapore), Vol. 17, pp. 80-3.  \\

\noindent Chaitin, G.J., 1999. {\it The Unknowable}, Springer
(Singapore), 122 pp. \\

\noindent Chambadal, L., 1981. {\it Dictionnaire de
math\'{e}matiques}, Hachette, Paris, pp. 225-26.  \\

\noindent Choquet, G, 1984. {\it Cours de Topologie}. Masson,
Paris, pp. 64-5. \\

\noindent Church, A., 1936. "An unsolvable problem of elementary
number theory", {\it Am. J. Math.}, Vol. 58, pp. 345-63.  \\

\noindent G\"{o}del, K., 1931. "On formally undecidable
propositions of Principia Mathematica and related systems I", {\it
Monatshefte f\"{u}r Mathematik und Physik}, Vol. 38, pp. 173-198.
\\

\noindent James, G., James, R.C., 1992. {\it Mathematics
Dictionary}, Van Nostrand Reinhold, New York, pp. 267-68.  \\

\noindent Klir, G.J., Wierman, M.J., 1999. "Uncertainty-based
information", {\it Studies in fuzziness and soft computing}, Vol.
15, Springer-Verlag, Berlin, New York, 168 p. \\

\noindent Krasnoholovets, V., Ivanovsky, D., 1993. "Motion of a
particle and the vacuum", {\it Physics Essays}, Vol. 6, No. 4, pp.
554-63. (Also arXiv.org e-print archive quant-ph/9910023).  \\

\noindent Krasnoholovets, V., 1997. "Motion of a relativistic
particle and the vacuum", {\it Physics Essays}, Vol. 10, No. 3,
pp. 407-16. (Also quant-ph/9903077).  \\

\noindent Krauss, L., 1999. "L'antigravit\'{e}", {\it Pour la
Science} ({\it Scientific American}, French ed.), Vol. 257, pp.
42-9. \\

\noindent Kubel, H., 1997. "The Lorentz transformation derived
from an absolute aether", {\it Physics Essays}, Vol. 10, No. 3,
pp. 510-23.  \\

\noindent Lin, Y., 1988. "Can the world be studied in the
viewpoint of systems?" {\it Math. Comput. Modeling,} Vol. 11, pp.
738-742. \\

\noindent Lin, Y., 1989. "A multirelational approach of general
systems and tests of applications. Sybthese". {\it Internal. J.
Epistemol. Methodol. and Philos. of Sci.}, Vol. 79, pp. 473-488.
\\

\noindent Malina, R., 2000. "Exploration of the invisible cosmos",
in: {\it Exploration: Art, Science \& Projects, Institute of
Ecotechnics Conference}, Aix-en-Provence, France, Oct. 27-30. \\

\noindent Meno, F., 1998. "A smaller Bang?" {\it Physics Essays},
Vol. 11, No. 2, pp. 307-10. \\

\noindent Mirimanoff, D., 1917. "Les antinomies de Russell et de
Burali-Forti, et le probl\`{e}me fondamental de la th\'{e}orie des
ensembles", {\it L'Enseignement Math\'{e}matique}, Vol. 19, pp.
37-52.  \\

\noindent Mitchell, W.C., 1995. {\it The Cult of the Big-Bang},
Cosmic Sense Books, Carson City, USA, 240 pp.  \\

\noindent Rothwarf, A., 1998. "An aether model of the universe".
{\it Physics Essays}, Vol. 11, No. 3, pp. 444-66. \\

\noindent Schwartz, L., 1991. {\it Analyse I: th\'{e}orie des
ensembles et topologie}, Hermann, Paris, pp. 30-5.  \\

\noindent Tricot, C., 1999. {\it Courbes et dimension fractale},
Springer-Verlag, Berlin Heidelberg, (See also: 1999a, pp. 240-60;
1999b: p.51, 110). \\

\noindent Turing, A.M., 1937. "On computable numbers, with an
application to the entscheidungsproblem", {\it Proc. Lond. Math.
Soc.}, Series 2, Vol. 42, pp. 230-65, and Vol. 43, pp. 544-6. \\

\noindent Weisstein, E.W., 1999. CRC Concise encyclopedia of
mathematics, Springer-Verlag, New York, (See also: 1999a, pp.
1099-1100; 1999b: pp. 473-474; 1999c: p. 740).  \\

\noindent Wesley, J. P., 1994. "Experimental results of Aspect et
al. confirm classical local causality", {\it Physics Essays}, Vol.
7, p. 240, (See also: 1998,  Vol. 11, No. 4, p. 610).  \\

\noindent Wesley, J. P., 1995. "Classical quantum theory", {\it
Apeiron}, Vol. 2, No. 2, pp. 27-32. \\

Wu, Y., Lin, Y., 2002. "Beyond nonstructural quantitative
analysis", in {\it Blown Ups, Spinning Currents and Modern
Science}, World Scientific, New Jersey, London, p. 324.  \\

\noindent Zadeh, L., Kacprzyk, J. (Eds.), 1992. {\it Fuzzy Logic
for the Management of Uncertainty}, John Wiley \& Sons, New York.
\\

\subsection*{Further reading}

Abbott, E.A., 1884. {\it Flatland: a romance of many dimensions},
publ. 1991 by Princeton University press, Princeton. \\

\noindent Dewdney, A.K., 2000. "The planiverse project: then and
now", {\it The Mathematical Intelligencer}, Vol. 22, No. 1, pp.
46-51. \\

\noindent Hannon R.J., 1998. "An alternative explanation of the
cosmological redshift", {\it Physics Essays}, Vol. 11, No. 4, pp.
576-578. \\

\noindent K\"{o}rtv\'{e}lyessy, L., 1999. {\it The electrical
universe}. Effo Kiado \'{e}s Nyomda, Budapest, 704 pp. \\

\noindent Krasnoholovets, V., Byckov, V., 2000. "Real inertons
against hypothetical gravitons. Experimental proof of the
existence of inertons". {\it Ind. J. Theor. Phys}., Vol. 48, No.
1, pp. 1-23. (Also quant-ph/0007027). \\

\noindent Krasnoholovets, V., 2000. "On the nature of spin,
inertia and gravity of a moving canonical particle", {\it Ind. J.
Theor. Phys.}, Vol. 48, No. 2, pp. 97-132. (Also
quant-ph/0103110). \\

\noindent Lebesgue, H., 1928. Le\c{c}ons sur 1'int\'{e}gration,
Ed. Coll\`{e}ge de France, Paris, p. 179. \\

\noindent Lester, J.,1998. "Does matter matter?" {\it Physics
Essays}, Vo. 11, No. 4, pp. 481-491.  \\

\noindent Loewenstein, W., 1999. {\it The touchstone of life:
molecular information, cell communication, and the foundations of
life}. Oxford University Press, Oxford, 368 pp.  \\

\noindent Meno, F., 1997. "The photon as an aether wave and its
quantized parameters", {\it Physics Essays}, Vol. 10, No. 2, pp.
304-314. \\

\noindent Verozub, L., 1995. "The relativity of spacetime", {\it
Physics Essays}, Vol. 8, No. 4, pp. 518-523.   \\

\noindent Watson, G., 1998. "Bell's theorem refuted: real physics
and philosophy for quantum mechanics", {\it Physics Essays}, Vol.
11 No. 3, pp. 413-421. \\

\subsection*{Appendix}

\hspace*{\parindent} A specimen case of calculation of the
dimension of a set whose members are put on the physicist's table
like detached parts.

Let ($\rm E^{{\kern 1pt}e}_{n}) = \{aa, {\kern 2pt}ba, {\kern
2pt}ca, {\kern 2pt}bc, {\kern 2pt}ab, {\kern 2pt}cc, {\kern
2pt}ac, {\kern 2pt}bb, {\kern 2pt}cb\}$. Applying (11.3a) gives
several results such as: $\rm \{a, {\kern 2pt}b, {\kern 2pt}bb,
{\kern 2pt}ca, {\kern 2pt}cc, {\kern 2pt}ab, {\kern 2pt}ac\}$ or
$\rm \{aa, {\kern 2pt}cc, {\kern 2pt}bb, {\kern 2pt}ca, {\kern
2pt}bc, {\kern 1pt}b, {\kern 2pt}a, \}$, etc. and $\rm \{aa,
{\kern 2pt}bb, {\kern 2pt}cc \}$. Thus: $\rm diam (E^{{\kern
1pt}e}_{{\kern 1pt}n}) \subseteq \{aa, {\kern 2pt}bb, {\kern
2pt}cc\}$ which matches with the diagonal of the Cartesian product
$\rm \{a,b,c\} \times \{a,b,c\}$. Then, applying (11.3b) gives
either \{\O\} or \{ab\}, or \{bc\}, or \{ac\}, that is cobbles
having one member of two nuclei as the max representing $\rm diag
(\Pi^{{\kern 1pt} e}_{{\kern 1pt} n})$. Since $\rm  diag
(E^{{\kern 1pt}e}_{{\kern 1pt}n})$ has three members of two nuclei
each, the size ratio is $\varrho = 1/3$ while the number of
cobbles tessellating the set is 9. Hence, applying either
equations (2.1)--(2.2) or equations (11.2) gives $\rm 9\cdot
(l/3)^{e} =1$, that is, $\rm e = \ln 9/\ln3 = 2$, or $\rm e = (\ln
9 + 2 \ln 2)/\ln 6 = 2$.

The dimension of ($\rm E^{1}_3)\times (E^{1}_3) = (E^{2}_3)$ has
been therefore correctly estimated. Had the set been alternatively
composed differently, as for instance: $\rm (E^{{\kern 1pt}
e}_{{\kern 1pt}n}) = \{aa, {\kern 2pt}ba, {\kern 2pt}ca, {\kern
2pt}bc, {\kern 2pt}ab, {\kern 2pt}cc, {\kern 2pt}ac, {\kern
2pt}bb\}$ having a lesser number of heterogeneous cobbles, then
one would have found: $\rm 8 \cdot (1/3)^{e} =1$, that is, $\rm e
= 1.89$, a noninteger dimensional exponent, indicating a space
with some fractal-like feature.

\end{document}